\documentclass[%
 aip,
 amsmath,amssymb,amsthm,
preprint,%
]{revtex4-1}

\usepackage{graphicx}
\usepackage{dcolumn}
\usepackage{bm}

\usepackage[utf8]{inputenc}
\usepackage[T1]{fontenc}
\usepackage{mathptmx}
\usepackage{color,float}
\usepackage{hyperref}

\renewcommand{\i}{\ensuremath{\text{\normalfont I}}}
\newcommand{\ii}{\ensuremath{\text{\normalfont I\!I}}}
\newcommand{\iii}{\ensuremath{\text{\normalfont I\!I\!I}}}
\newcommand{\iiii}{\ensuremath{\text{\normalfont I\!V}}}

\let\dsp=\displaystyle

\def\eg{{\it e.g.}}
\def\ie{{\it i.e.}}

\begin{document}

\preprint{AIP/123-QED}

\title[Non-Markovian Schr\"{o}dinger ]{ Markovian Embedding Procedures for  Non-Markovian Stochastic Schr\"{o}dinger Equations}

\author{Xiantao Li}%
 \email{Xiantao.Li@psu.edu.}
\affiliation{ 
The Pennsylvania State University, University Park, Pennsylvania, 16802, USA.
}%

\date{\today}

\begin{abstract}
We present embedding procedures for  the non-Markovian stochastic Schr\"{o}dinger equations, arising from studies of quantum systems coupled with bath environments. 
By introducing auxiliary wave functions, it is demonstrated that the non-Markovian dynamics can be embedded in extended, but Markovian, stochastic models. Two embedding procedures are presented. The first method leads to nonlinear stochastic equations, the implementation of which is much more efficient than the non-Markovian stochastic Schr\"{o}dinger equations. 
 The stochastic  Schr\"{o}dinger equations obtained from the second procedure involve more auxiliary wave functions, but the equations are linear, and we derive the corresponding generalized quantum master equation for the density-matrix. 
The accuracy of the embedded  models is ensured by fitting to the power spectrum.  The stochastic force is represented using a linear superposition of Ornstein–Uhlenbeck processes, which are incorporated as multiplicative noise in the auxiliary Schr\"{o}dinger equations.
The asymptotic behavior of the spectral density in the low frequency regime is preserved  by using correlated stochastic processes.  
 The approximations are verified by using a spin-boson system as a test example.
 \end{abstract}

\maketitle

\section{\label{intro} Introduction}

A fundamental problem in quantum chemistry is the dynamics of a quantum system (S) interacting continuously with its bath (B) environments. One outstanding challenge is the description of non-Markovian dynamics, which arises when there is a lack of separation of time scales between the system and bath. The interest in studying such scenario is further driven by overwhelming experimental evidence  suggesting  many processes  that are highly non-Markovian \cite{groblacher2015observation,madsen2011observation}. Often observed as  memory effects, the non-Markovian nature might be tied to important quantum properties, such as quantum decoherence, correlations and entanglement \cite{breuer_non-markovian_2016,shiokawa1995decoherence,bellomo2007non,thorwart2009enhanced}.
 
Theoretically the non-Markovianity can be made more precise and further quantified by using appropriate criteria  \cite{breuer_non-markovian_2016}. On the other hand, an accurate and practical description is highly challenging since a direct simulation of the combined system with degrees of freedom in the bath explicitly represented is not yet feasible. Much effort has been focused on deriving a reduced model, where only the degrees of freedom in the system are resolved. Existing methods can roughly be divided into two categories \cite{de_vega_dynamics_2017}. The first type of approaches  accomplish the reduction of computational complexity  based on the density-matrix, using  the Nakajima-Zwanzig (NZ) projection formalism \cite{nakajima1958quantum,zwanzig1960ensemble} for the Liouville von Neumann equation,  to derive an equation for the reduced density-matrix (RDM)   $\rho_S$ of the system.  The projection operator is defined using the trace operator over the bath variables, and the reduced equation involves a memory integral that naturally describes the non-Markovian dynamics. In principle, the density-matrix equation derived this way is exact. But the memory superoperator is too abstract to be implemented directly. Various approximations have since  been proposed to simplify the NZ equation 
  \cite{esposito2003quantum,cao_lindblad_2017,van1992stochastic,shi_new_2003}.  The simplest approximation, known as the Born-Markov approximation, leads to a Markovian dynamics, for which as proved by Lindblad \cite{lindblad1976generators},  the generator can be completely characterized. The Lindblad equation is also known as the quantum master equation (QME) and it has served as an important description for open quantum systems.    Meanwhile, a lot of effort has been made to model the non-Markovian dynamics faithfully, using more accurate approximations of the memory operator \cite{pfalzgraff_efficient_2019,meier_non-markovian_1999}.  The resulting density-matrix equation has been regarded as generalized quantum master equation (GQME). A more recent approach, also in a similar spirit, employs the Mori's projection \cite{Mori65}, which is defined using inner products \cite{kelly2016generalized,montoya2017approximate,montoya2016approximate}.      Chru\'scin\'ski and Kossakowski  \cite{chruscinski_non-markovian_2010} showed how the non-Markovian dynamics can be described by a local dynamics with a non-Markovian generator. More recently, Banchi  et al. \cite{banchi2018modelling}  presented a data driven approach, where the memory effect is modelled through a recurrent neural network.

The other type of approaches work with the wave functions and model the dynamics of the quantum system using non-Markovian stochastic Schr\"odinger equations (SSE).  
Using perturbation techniques, Gaspard and Nagaoka \cite{de_vega_non-markovian_2005,gaspard1999non} derived SSEs that involve a memory integral and correlated Gaussian noise.   In terms of the coupling constant $ \lambda$, the non-Markovian SSEs  by Gaspard and Nagaoka retain terms up to  $\mathcal{O}(\lambda^3)  $ order. 
 Compared to the density-matrix, the choice of the initial conditions for wave functions is more involved \cite{hortikar1998correlations,srednicki1994chaos}. On the other hand, the memory term is much simpler than that in the GQME.  
 Another important approach to derive non-Markovian SSEs is by using path integrals and coherent states  \cite{diosi1996exact,diosi1997non,strunz1996linear}, which is in principle exact. An important difference from those SSEs obtained by Gaspard and Nagaoka is that the memory integral is a time convolution of the functional derivative of the wave function with respect to a underlying Gaussian noise. 
Later, these non-Markovian SSEs were simplified by introducing an hierarchy of pure states (HOPS)  
 \cite{suess_hierarchy_2014}.  
 By approximating the memory term with a sum of exponentials and introducing auxiliary pure states, the memory term can be embedded in an extended system, from which the density matrix can be computed as an ensemble average.  Ke and Zhao also proposed a hierarchy of forward--backward SSEs 
 as a stochastic unravelling of the  functional in the path-integral formalism of reduced density operator \cite{ke2016hierarchy}.

The starting points of both types of approaches are in principle consistent.
In the case when there is sufficient separation between the time scales of the system and bath, a Markovian approximation can be introduced in both approaches, and the reduced models at this level are also consistent \cite{gisin1992quantum}. More specifically, the density-matrix approach would yield the Lindblad equation. On the other hand, the wave function based approach, with the random force approximated by white noise,  leads to a set of stochastic differential equations (SDEs), for which the second moment satisfies the Lindblad equation. But in the modeling of non-Markovian dynamics,  such an exact connection breaks down. For instance, the non-Markovian SSEs are stochastic integro-differential equations, and in general, there is no closed-form equation for the density-matrix. For example, the GQMEs derived by Gaspard and Nagaoka \cite{gaspard1999non,gaspard1999slippage,esposito2003quantum}  holds only approximately, up to $\mathcal{O}(\lambda^3) $ terms.  A more subtle situation is where the density-matrix from some of the non-Markovian models  can not be guaranteed to be positive \cite{van1992stochastic}. Therefore, they can not be unravelled by   SSEs.

The derivations of the non-Markovian SSEs  \cite{gaspard1999non,diosi1997non,strunz_open_1999} point to an important relation between the random force and the memory operator. This bears a remarkable analogy to the second fluctuation dissipation theorem (FDT) in the generalized Langevin equations (GLE) in classical mechanics \cite{Kubo66}. This relation will hereafter be referred to as a {\it statistical consistency}. 
This paper presents approximations of the non-Markovian SSE derived  by  Gaspard and Nagaoka \cite{gaspard1999non}. Using complex-valued Ornstein–Uhlenbeck (OU) processes as building blocks, we construct approximations of the random force with coefficients determined by a least-square fitting of the power spectrum.  The aforementioned  statistical consistency leads to  an approximation of the memory kernel in the form of a sum of exponentials. In this case, the memory integral can be embedded in an extended, but Markovian dynamics, by introducing auxiliary wave functions. Compared to a direct  treatment of the non-Markovian SSE, the implementation of the extended system is considerably more efficient.

 The idea of using auxiliary wave functions to encode the memory effect has also been pursued by Suess et al. \cite{suess_hierarchy_2014}  to simplify the non-Markovian  SSEs derived from the path integral approach  \cite{strunz_open_1999}.  Truncations are made to the hierarchy to obtain a finite system. The current approach starts with the SSEs  derived  by  Gaspard and Nagaoka \cite{gaspard1999non}. The superposition of OU processes are used as an efficient method to approximate the noise. We show that the asymptotic behavior of the spectral density can be preserved  constructing  correlated OU processes. 
This paper will also introduce another type of embedding, where the noise is blended into additional auxiliary wave functions. In this case, the corresponding extended SSEs are {\it linear}. As a result, unlike the first type of embedding, a closed-form GQME can be derived.  The GQME derived from this embedding procedure is different from those derived by Meier and Tannor  \cite{meier_non-markovian_1999} and Tanimura \cite{tanimura2006stochastic}, where a similar approximation was made to the spectral density, and auxiliary density-matrices were introduced to encode the memory effect. Such auxiliary density matrices all have the same dimension as the original density matrix. When compared with the extended density matrix from our embedding procedure, there is no off-diagonal blocks.

The two types of embeddings presented here are two descriptions that are complementary. In the case when the wave functions for the quantum system  is extended in space, \eg, those applications in molecular junctions, simulating the SSEs will likely be more efficient than the GQME \cite{gisin1992quantum}. On the other hand, when the quantum system involves a small number of states, \eg, spins or qubits \cite{shiokawa1995decoherence}, the GQME  approach would be more efficient since no ensemble average is needed.  
Aside from the practical implementations, we would like to point out that there is not much theory for non-Markovian stochastic processes (one exception is  stationary Gaussian processes \cite{Doob44}). In contrast, Markovian dynamics has been a very mature subject. A wide variety of   analytical theory  \cite{oksendal2003stochastic}, computational techniques  \cite{kloeden2013numerical},   and data-driven statistical methods \cite{iacus2009simulation} are available. Therefore, the author believes that the extended SSEs, with correctly calibrated parameters, are better suited as  descriptions of non-Markovian quantum dynamics.

\section{\label{th} Theory and algorithms}

\subsection{Non-Markovian Stochastic Schr\"{o}dinger Equations}

We consider a quantum system-bath model with the standard setup, where the total Hamiltonian is expressed using  tensor products of
operators on the Hilbert spaces associated with the system and bath,
\begin{equation}\label{eq: htot}
 H_{tot}= H_S \otimes I_B + I_S \otimes H_B + \lambda \sum_{j=1}^J S_j \otimes B_j.
\end{equation}
Here $\lambda$ is a coupling constant, and we make the weak coupling assumption, \ie, $0< | \lambda  |\ll 1.$

To study the non-Markovian dynamics of the system variables, we consider the approach by Gaspard and Nagaoka \cite{gaspard1999non}, where a perturbation technique is employed and the following non-Markovian stochastic Schr\"{o}dinger equation is derived,
\begin{equation}\label{eq: sse}
    i \partial_t \psi = \hat H_S \psi - i \lambda^2 \int_0^t c(\tau) S
    e^{-i \hat H_S \tau } S \psi(t-\tau) d\tau + \lambda  \eta(t)  S \psi(t).
\end{equation}
Here $i=\sqrt{-1}.$ For simplicity, we have chosen the number of baths to be 1. 

It has also been shown by Gaspard and Nagaoka \cite{gaspard1999non} that the noise term $\eta(t)$ is Gaussian with mean zero and correlation given by,
\begin{equation}\label{eq: ct}
    \overline{\eta(t)^* \eta(t') }= c(t-t'). 
\end{equation}
The stationarity of the process also implies that $c(t)=c(-t)^*$, and so it suffices to show the correlation for $t\ge 0.$ The overline indicates the statistical average with respect to the equilibrium Gaussian distribution. This relation between the random noise and the kernel function in the integral is reminiscent of the second fluctuation-dissipation theorem \cite{Kubo66} in classical statistical mechanics, and it has been a primary motivation of the current work.  But it is also worthwhile to point out  the important difference that the noise in \eqref{eq: sse} is of multiplicative nature. 

Due to the memory term, and the time correlation of the noise, the dynamics associated with the SSE \eqref{eq: sse} is non-Markovian. The relation \eqref{eq: ct} suggests that approximations  of those two terms should be done simultaneously in a consistent manner.
 The simplest approximation  is by a white noise, $ \eta(t) =  \dot W(t)$, where $W(t)$ is the standard complex-valued Brownian motion. In this case,  the correlation is reduced to a delta function, 
\begin{equation}\label{eq: fdt}
    c(t-t')= \gamma^2 \delta (t-t').
\end{equation}

With this approximation, the SSE \eqref{eq: sse} becomes  Markovian,
\begin{equation}\label{eq: sse-mark}
    \partial_t \psi = -i\hat H_S \psi - \frac{i}2 \delta^2  S^\dagger S \psi(t) + \delta S \psi(t) \dot{W}(t).   
\end{equation}
We have combined the coefficients by letting $\delta =\gamma \lambda$. This model was referred to as $\delta$-correlated bath by Gaspard and Nagaoka \cite{gaspard1999non}, and it  has been used in the study the electron transport problems \cite{biele2012stochastic}, and integrated with the time-dependent density-functional theory \cite{di2007stochastic}. An important property \cite{gisin1992quantum} is that the density-matrix, 
$\rho = \overline{\psi \psi^*}$,
satisfies the quantum master equation (QME) known as the Lindblad equation \cite{lindblad1976generators},
\begin{equation}
    i \partial_t \rho  =[H, \rho] -\frac{i}{2}({S}^\dagger {S}\rho + \rho {S}^\dagger {S}-2{S}\rho {S}^\dagger). 
\end{equation}

\subsection{Markovian embedding using auxiliary orbitals}
In this paper we aim to extend the $\delta$-correlated bath approximation to correlated processes. To illustrate the ideas, let us first begin with a complex-valued Ornstein–Uhlenbeck (OU) process \cite{risken1984fokker},
\begin{equation}\label{eq: cou}
    i \dot{\zeta} = -\alpha \zeta +  \gamma \dot{W}(t). 
\end{equation}
Here, the $\dot{\;}$ indicates the time derivative, and $\alpha$ is a complex number with  $\text{Im} \alpha \ge 0$ to ensure stability.  Furthermore, $\gamma$ is a real number, and $W(t)$ is the standard complex-valued Wiener process. We choose the initial condition to be Gaussian with variance one, \ie,
\begin{equation}\label{eq: eta0}
\overline{\zeta(0)}=0, \;\;\overline{\zeta(0)^* \zeta(0)} = 1. 
\end{equation}
It can be shown that when
\begin{equation}\label{eq: fdt0}
 \gamma^2= 2\text{Im} (\alpha),
\end{equation}
 $\zeta(t)$ is stationary with correlation 
\begin{equation}\label{eq: 1exp}
    \overline{\zeta(t)^* \zeta(t') } =  e^{-i\alpha^* (t-t')}, \quad t\ge t'.
\end{equation}
We have left the proof in the Appendix.  The relation \eqref{eq: fdt0} reflects the fluctuation-dissipation theorem \cite{Kubo66}. 

Such OU processes will be used as building blocks to generate Gaussian processes with more general correlation functions, since in practice, the correlation function $c(t)$ can often by efficiently approximated by a short sum of complex exponentials \cite{ritschel2014analytic} (We will discuss this aspect later). Consequently, both the memory integral and the correlated Gaussian process $\eta(t)$ can be simplified considerably. 

To further elaborate this idea,
let us make the approximation $c(t) \approx \theta^2 e^{-i\alpha^* t} $, and insert it into the memory term in SSE \eqref{eq: sse}. Letting, 
\[ \chi(t) =  \lambda \theta^{-1} \int_0^t c(\tau) 
    e^{-i \hat H_S \tau } S \psi(t-\tau) d\tau, \quad \chi(0)=0, \]
and by choosing $\eta(t)$ in accordance with \eqref{eq: fdt},
\begin{equation}
\eta(t) \approx \theta \zeta(t),
\end{equation}
we obtain from \eqref{eq: sse} that,
\begin{equation}\label{eq: psi}
     i \partial_t \psi = \hat H_S \psi - i \lambda \theta S^\dagger \chi + \lambda \theta \zeta(t) S \psi(t).
\end{equation}

 With direct differentiation of $\chi$, we find
\begin{equation}\label{eq: ch1}
    i\partial_t \chi = (H_S + \alpha^* ) \chi + i \theta \lambda S \psi(t).  
\end{equation}
As a result, the memory term is completely incorporated into this auxiliary equation. By collecting equations, we arrive at the following closed-form  stochastic differential equations (SDEs),
\begin{equation}\label{eq: sse-I}
\left\{\begin{aligned}
   i \partial_t \psi =& \hat H_S \psi - i \lambda \theta S^\dagger \chi + \lambda \theta \zeta(t) S \psi(t),\\
   i\partial_t \chi = & (H_S + \alpha^* ) \chi + i \theta \lambda S \psi(t), \\
      i \dot{\zeta} = &-\alpha \zeta +  \gamma \dot{W}(t). 
\end{aligned}\right.
\end{equation}
We will refer to this derivation of a stochastic system as type {\bf I} embedding. The function $\chi(t)$ is an auxiliary wave function. This procedure yields an extended, but {\it Markovian}, stochastic dynamics, for which a variety of analytical, computational and statistical methods are available.  In the case when the correlation function $c(t)$ is approximated by a sum of exponential functions, this embedding can be applied to each mode. This will be discussed in the next section.   

\medskip

When the simulations are based on the wave function via SSEs \eqref{eq: sse}, the type {\bf I} embedded model \eqref{eq: sse-I} offers an efficient alternative, since the history of the wave functions does not need to be stored, and the numerical integration at each step is replaced by a one-step integration of  $\chi$, which is also much cheaper. 
To give a perspective, at the $n$th time step, the approximation of the integral in \eqref{eq: sse} requires $\mathcal{O} (n) $ operations. Therefore, evolving the SSE  \eqref{eq: sse} for $N$ steps would require $\mathcal{O}(N^2)$ operations.  In contrast,  at each step,  the embedded model \eqref{eq: sse-I} only requires solving the extra equation for one step. Therefore, the total computation is still $\mathcal{O}(N)$.

On the other hand, there are cases where simulations based on the density-matrix is more feasible, e.g., when the system consists of a few spins. However, due to the nonlinearity in the second equation of   the type {\bf I} embedding \eqref{eq: sse-I}, typically there is no closed-form for the corresponding density-matrix equation.  To address this issue, we extend our embedding procedure as follows: We write $\chi^\i= \chi, $ and define another auxiliary wave function,
\begin{equation}\label{eq: chi2}
\chi^\ii= i  \zeta(t) \psi(t),
\end{equation}
 to incorporate the noise as well.  By It\^{o}'s lemma, we have,
\begin{equation}\label{eq: chi2'}
    i\partial_t \chi^\ii = (\hat H_S - \alpha) \chi^\ii + i  \gamma \dot{W} \psi(t) + \lambda \theta \zeta(t) S^\dagger \chi^\i +  \lambda\theta \zeta(t) 
    S\chi^\ii. 
\end{equation}

At this point, we will drop the last two terms on the right hand side. These two terms, when substituted back into \eqref{eq: psi}, will contribute to an $\mathcal{O}(\lambda^2)$ term in the error. With this truncation, we can collect the equations,
\begin{equation}\label{eq: ext0}
\left\{
    \begin{aligned}
       i \partial_t \psi =& \hat H_S \psi - i \lambda \theta  S^\dagger \chi^\i - i \lambda \theta S \chi^\ii, \\
       i\partial_t \chi^\i =& (\hat H_S + \alpha^* ) \chi^\i + i \theta \lambda S \psi(t),\\
       i\partial_t \chi^\ii = & (\hat H_S - \alpha) \chi^\ii  + i \gamma \psi(t) \dot{W} .
    \end{aligned} \right.
\end{equation}

As for the initial conditions, we have
\begin{equation}\label{eq: initchi}
 \chi^\i(0)=0,  \text{and} \; \chi^\ii(0)= \xi \psi(0),
\end{equation}
 with $\xi$ being a complex Gaussian random variable with mean zero and variance $1$, as can be seen from \eqref{eq: chi2} and \eqref{eq: eta0}.  

This procedure will be called type {\bf II}  embedding and the resulting SDEs will be referred to as type {\bf II} embedded SSEs.  Similar to type {\bf I} embedding, this procedure obtains an extended Markovian system.  The main departure from type {\bf I} embedding is that the effect of the noise is embedded in another SSE using a second auxiliary wave function $\chi^\ii$.  The advantage is that the nonlinear term in  \eqref{eq: sse-I}  has been eliminated. As a result, one can write down the density-matrix equation in a closed form, which then enables  a generalized quantum master equation (GQME) model. This will be presented in section {\bf II.} E. 

We also point out that one can add a $ i \theta \lambda S^\dagger \psi(t)$ term  in the equation for $\chi^\ii$, as a conjugate pair  to the last term in the first equation. This modification is still within an $\mathcal{O}(\lambda^2)$ approximation of the original SSE. But it makes  the entire stochastic system stable. This can be directly verified by examining the skew-Hermitian part of the Hamiltonian operator in \eqref{eq: ext0}. 

%

The  type {\bf II}  embedding  procedure can be continued to obtain a higher order embedding in terms of the coupling constant $\lambda$. Toward this end, we keep Eq. \eqref{eq: chi2'}, and further define,
\begin{equation}\label{eq: 34}
  \chi^\iii = i\zeta(t) \chi^\i, \quad \phi^\iiii= i \zeta(t) \chi^\ii,
\end{equation}
which turns Eq. \eqref{eq: chi2'} into,
\begin{equation}\label{eq: chi2''}
\begin{aligned}
    i\partial_t \chi^\ii =& (\hat H_S - \alpha) \chi^\ii- i \lambda \theta   S^\dagger \chi^\iii - i \lambda  \theta 
    S\chi^\iiii  + i \gamma \dot{W} \psi(t) , \\
     i\partial_t \chi^\iii =& i\lambda \theta S \chi^\ii +  (\hat H_S - \alpha + \alpha^*) \chi^\iii +i {\gamma} \chi^\i \dot{W}, \\
         i\partial_t \chi^\iiii =&   i\lambda \theta S^\dagger \chi^\ii + (\hat H_S - 2 \alpha ) \chi^\iiii  + 2i  {\gamma}\chi^\ii \dot{W}. \\
\end{aligned}
\end{equation}
Here we dropped $\mathcal{O}(\lambda)$ terms in the last equation. When substituted back into Eq. \eqref{eq: psi}, they only contribute to   $\mathcal{O}(\lambda^3)$ terms.
The initial conditions follow from Eq. \eqref{eq: 34},
\[ \chi^\iii(0)=0, \quad \chi^\iiii(0)= -{\zeta(0)^2} \psi(0).\]

\subsection{Approximation of the correlation function by a sum of exponentials}

In many cases the function $c(t)$ in \eqref{eq: ct} can be well approximated by a sum of exponential functions \cite{ritschel2014analytic},
\begin{equation}\label{eq: sum-exp}
    c(t)\approx \sum_{k=1}^{k_\text{max}} \theta_k^2 e^{- i \alpha_k^* t}, \quad \alpha_k = \mu_k + i \nu_k.
\end{equation}
In this approximation it is enough to assume $t\ge 0$, and for $t<0$  the values of the correlation function can be obtained from the relation  $c(-t) = c(t)^*$. Here we have chosen the weights $\theta_k^2$ to be positive. Thus, each of the terms can be regarded as the correlation of an OU process. Namely,
$$ e^{-\alpha_k^* t} = \overline{\zeta_k(t)^* \zeta_k(0)},$$ where $\zeta_k(t)$ follows the SDEs with independent Brownian motions $W_k(t)$,
\begin{equation}\label{eq: cou'}
    i \dot{\zeta}_k = -\alpha_k \zeta_k +  \sqrt{2 \nu_k} \dot{W}_k(t). 
\end{equation}
We have kept the relation \eqref{eq: fdt0}. The random noise $\eta(t)$ in \eqref{eq: sse} can be generated by,
\begin{equation}\label{eq: zeta2eta0}
 \eta(t) \approx  \sum_{k=1}^{k_\text{max}} \theta_k \zeta_k(t).
\end{equation}

To look at the approximation \eqref{eq: sum-exp} in the frequency domain, we take the Fourier transform, 
\begin{equation}
|G(\omega)|^2  = \mathcal{F}[c(t)]= \int_{-\infty}^{+\infty} e^{-i\omega t} c(t) dt,
\end{equation}
which corresponds to the power spectrum of the noise $\eta(t).$

Applying Fourier transform to both sides of  \eqref{eq: sum-exp}, we arrive at,
\begin{equation}\label{eq: Gw}
|G(\omega)|^2 \approx  \sum_{k=1}^{k_\text{max}} \frac{2 \theta_k^2 \nu_k}{(\omega + \mu_k)^2 + \nu_k^2 }.
\end{equation}
Therefore, the power spectrum is approximated by a sum of Lorentzians. The parameters can be obtained from a fitting procedure, \eg, a optimization method using simulation annealing \cite{meier_non-markovian_1999}.

\medskip

Now we return to the type {\bf I} embedding of the SSE \eqref{eq: sse}. By repeating  the embedding procedure from the previous section, we obtain an extended system  as follows,
\begin{equation}\label{eq: sse-I'}
\left\{
    \begin{array}{l}
     \; \; i \partial_t \psi =\dsp\;\; \hat H_S \psi - i \lambda  S^\dagger \sum_k \theta_k\chi_k + \lambda \sum_k \theta_k \zeta_k(t) S \psi(t),\\
       \begin{array}{rl}
           i\partial_t \chi_k = & (H_S + \alpha^*_k ) \chi_k + i \theta_k \lambda S \psi(t), \\
      i \dot{\zeta}_k = &-\alpha_k \zeta_k +  \gamma_k \dot{W}_k(t). 
       \end{array}
\qquad       k=1,2,\cdots, k_\text{max}.
    \end{array} \right.
\end{equation}

Similarly, we can extend the type {\bf II} embedded models to incorporate multiple OU processes,

\begin{equation}\label{eq: ext-ii}
\left\{
    \begin{array}{l}
        i \partial_t \psi =\dsp  \hat H_S \psi - i \lambda   S^\dagger \sum_k \theta_k \chi^\i_k - i \lambda  S \sum_k \theta_k \chi^\ii_k, \\
    \begin{array}{ll}
            i\partial_t \chi^\i_k =& (\hat H_S + \alpha^*_k ) \chi^\i_k + i \theta_k \lambda S \psi(t),\\
       i\partial_t \chi^\ii_k = & (\hat H_S - \alpha_k) \chi^\ii_k  + i  \gamma_k \psi(t) \dot{W}_k.
    \end{array}
    \qquad       k=1,2,\cdots, k_\text{max}.\end{array}
     \right.
\end{equation}

\subsection{Embedding with correlated OU processes}

The previous methods use linear superpositions of {\it independent} OU processes to approximate the random noise $\eta(t). $ The stability requires 
$\nu_k \ge  0$. Therefore, the power spectrum in Eq. \eqref{eq: Gw} is approximated by a sum of Lorenzians with {\it positive} coefficients.  This implies that 
the approximate power spectrum approaches to a finite constant near $\omega =0.$ { This is often inconsistent with the low frequency asymptotics of the spectral density of the bath.}

As an example, we consider the case when $|G(\omega)|^2 = \mathcal{O}(\omega^2)$ near the origin. Other asymptotic behavior can be similarly treated. We replace the approximation \eqref{eq: Gw} by the following ansatz,
\begin{equation}\label{eq: Gwc}
  |G(\omega)|^2 \approx  \sum_{k=1}^{k_\text{max}/2} Q_k L_k(\omega), \quad L_k(\omega) =   \frac{  \omega^2}{\big[ (\omega + \mu_{2k-1})^2 + \nu_{2k-1}^2 \big] \big[(\omega + \mu_{2k})^2 + \nu_{2k}^2 \big]  }.
\end{equation}
This is a slight generalization of symmetric or anti-symmetrized Lorentzians that were considered in \cite{meier_non-markovian_1999,ritschel2014analytic}.

We find that the corresponding noise can be realized by using two correlated OU processes as follows,
\begin{equation}
\left\{\begin{aligned}
  i \dot \zeta_{2k-1} =& - \alpha_{2k-1} \zeta_{2k-1} + \gamma_{2k-1} \dot{W}(t), \\
  i \dot \zeta_{2k} =& - \alpha_{2k} \zeta_{2k} + \gamma_{2k} \dot{W}(t), \\
    \alpha_{2k-1} =& \mu_{2k-1} + i \nu_{2k-1}, \quad \alpha_{2k} = \mu_{2k} + i \nu_{2k},\\
    \gamma_{2k-1} =& \sqrt{2\nu_{2k-1}},  \quad   \gamma_{2k} = \sqrt{2\nu_{2k}}.
\end{aligned}\right.
\end{equation} 
Notice that these two SDEs are driven by the {\it same} white noise $\dot W(t)$. The stationarity of the processes is ensured by
the last equation. Furthermore, the equilibrium covariance of the processes $(\zeta_{2k-1}, \zeta_{2k})$ is given by,
\begin{equation}\label{eq: cov}
\overline{
\left(
\begin{array}{c}
\zeta_{2k-1}^*\\
 \zeta_{2k}^*
 \end{array} \right)
 \left(
 \begin{array}{cc}
  \zeta_{2k-1} &\zeta_{2k} 
  \end{array} \right)}
  = 
\left(  \begin{array}{cc}
1 & Q_{12} \\
 Q_{12}^* & 1 
 \end{array}\right ), \quad  Q_{12}=  \frac{\gamma_1 \gamma_2}{\alpha_2 -\alpha_1^*}. 
\end{equation}
The calculations that led to these formulas can be found in the Appendix B. To sample these two Gaussian random variables, we can use the Cholesky's factorization of the above matrix as follows. Starting with two independent Gaussian random variables $z_1$ and $z_2$, we  define,
\begin{equation}\label{eq: samp}
 \zeta_{2k-1} = z_1, \quad  \zeta_{2k} =  Q_{12} \;z_1 + \sqrt{1- |Q_{12}|^2} \;z_2. 
\end{equation}

One can show that a linear combination of the two OU processes has  power spectrum given exactly be $L_k(\omega)$. Namely,
\begin{equation}
\mathcal{F}\big[  \overline{\zeta^*(t) \zeta(0) } \big] = Q_kL_k(\omega),
\end{equation}
where $\mathcal{F}$ denotes the Fourier transform, and $\zeta = c_{2k-1} \zeta_{2k-1} + c_{2k} \zeta_{2k}$ with coefficients given by,
\begin{equation}\label{eq: c12}
  c_{2k-1}= -\frac{Q_k\alpha_{2k-1}}{ (\alpha_{2k}-\alpha_{2k-1})\gamma_{2k-1}}, \quad  c_{2k}=\frac{Q_k\alpha_{2k}}{ (\alpha_{2k}-\alpha_{2k-1})\gamma_{2k}}.
\end{equation}

By Cauchy's residue Theorem, one can identify the corresponding correlation function in the time domain,
\begin{equation}\label{eq: torc}
\begin{aligned}
 \mathcal{F}^{-1}\big[Q_k L_k(\omega)\big]=& \theta_{2k-1}^2 e^{-i \alpha_{2k-1}^* t} + \theta_{2k}^2\;e^{-i \alpha_{2k}^* t}, \\
 \theta_{2k-1}^2 = & \frac{Q_k {\alpha_{2k-1}^*}^2 }{2\gamma_{2k-1}(\alpha_{2k}-\alpha_{2k-1}^*)(\alpha_{2k}^*-\alpha_{2k-1}^*)}, \\
 \theta_{2k}^2 = &\frac{Q_k {\alpha_{2k}^*}^2 }{2\gamma_{2k}(\alpha_{2k}^*-\alpha_{2k-1})(\alpha_{2k}^*-\alpha_{2k-1}^*)} .
\end{aligned}
\end{equation}
Therefore, the approximation \eqref{eq: Gwc} also corresponds to a sum of exponentials in the time domain. But compared to \eqref{eq: Gw}, the coefficients are allowed to be complex. 

The coefficients in the approximation \eqref{eq: Gwc} can be determined from fitting as well. The embedding of the SSEs using these correlated OU processes is surprisingly straightforward: In the SDEs \eqref{eq: sse-I'}, we simply let the white noise $W_{2k}$ to be the same as $W_{2k-1}$. In addition, the coefficients $\theta_k$'s in the first equation are replaced by $c_k$'s. Namely,  
\begin{equation}\label{eq: ck2eta}
 \eta(t)  \approx \sum_{k} c_k \zeta_k(t).
\end{equation}
In addition, the time correlation is being approximated as,
\begin{equation}\label{eq: torc-all}
 c(t)\approx \sum_{k} \theta_k^2  e^{-i\alpha_k^* t}. 
\end{equation}
This leads to the following type {\bf I} embedding,
\begin{equation}\label{eq: sse-I''}
\left\{
    \begin{array}{l}
       i \partial_t \psi =\dsp \hat H_S \psi - i \lambda  S^\dagger \sum_k \theta_k\chi_k + \lambda \sum_{k=1}^{k_\text{max}} c_k \zeta_k(t) S \psi(t),\\
           i\partial_t \chi_k =  (H_S + \alpha^*_k ) \chi_k + i \theta_k \lambda S \psi(t), \qquad k=1,2,\cdots, k_\text{max} \\
       \begin{array}{ll}
      i \partial_t{\zeta}_{2k-1} &= -\alpha_{2k-1} \zeta_{2k-1}+  \gamma_{2k-1} \dot{W}_k(t),\\
      i \partial_t{\zeta}_{2k} &= -\alpha_{2k} \zeta_{2k }+  \gamma_{2k} \dot{W}_k(t),
       \end{array}
   \qquad        k=1,2,\cdots, k_\text{max}/2. 
    \end{array} \right.
\end{equation}

Formally, the auxiliary equations can be solved and substituted into \eqref{eq: psi}, which leads to an approximation of the non-Markovian SSE \eqref{eq: sse}. This approximation exactly preserves the relation \eqref{eq: ct}.

For the type {\bf II} embedded models, we have,
\begin{equation}\label{eq: sse-II''}
\left\{
    \begin{array}{l}
       i \partial_t \psi =\dsp \hat H_S \psi - i \lambda  S^\dagger \sum_k \theta_k\chi_k^\i + \lambda \sum_{k=1}^{k_\text{max}} c_k S\chi^\ii_k(t),\\
           i\partial_t \chi_k^\i =  (H_S + \alpha^*_k ) \chi_k^\i + i \theta_k \lambda S \psi(t), \qquad k=1,2,\cdots, k_\text{max} \\
       \begin{array}{ll}
      i \partial_t{\chi}_{2k-1}^\ii &= (\hat H_S -\alpha_{2k-1}) \chi^\ii_{2k-1}+  i\gamma_{2k-1} \dot{W}_k(t),\\
      i \partial_t{\chi}_{2k}^\ii &= (\hat H_S-\alpha_{2k}) \chi^\ii_{2k }+ i \gamma_{2k} \dot{W}_k(t),
       \end{array}
   \qquad        k=1,2,\cdots, k_\text{max}/2. 
    \end{array} \right.
\end{equation}
The initial conditions are given by, ${\chi}_{2k-1}^\ii(0) =  i \zeta_{2k-1}(0) \psi$ and ${\chi}_{2k}^\ii(0) =  i \zeta_{2k}(0) \psi$. $\zeta_{2k-1}(0)$ and $\zeta_{2k}(0)$ can be sampled  using the covariance matrix \eqref{eq: cov} and the matrix factorization \eqref{eq: samp}.

\subsection{The generalized quantum master equation for type {\bf II} embedding} 
To derive the corresponding GQME 
of the stochastic model \eqref{eq: ext0}, we first write it  as a system of SDEs,
\begin{equation}\label{eq: Psi-sde}
  i \partial_t \Psi = H \Psi + \sum_k V_k \psi  \dot{W}_k.
\end{equation}
Here the function $\Psi$ includes the wave function $\psi$ and the auxiliary wave functions $\{\chi_k^\i, \chi_k^\ii\}_k$. It is often convenient to write $H=H_0+H_1$, with $H_0$ being the Hermitian part of the Hamiltonian. $\Psi$ and $H_0$ can be expressed in a blockwise form,
\begin{equation}\label{eq: Psi}
  \Psi = \left(
  \begin{array}{c}
    \psi \\
    \chi_1^\i\\
    \chi_1^\ii\\
    \chi_2^\i\\
    \chi_2^\ii\\
   \vdots 
  \end{array}
  \right), \quad 
 H_0= \left[\begin{array}{cccccc}
 H_s & -i \lambda \theta_1  S^\dagger  & -i \lambda \theta_1  S  & -i \lambda \theta_2 S^\dagger &  -i \lambda \theta_2 S & \cdots  \\
 i \lambda \theta_1 S & H_s + \mu_1 & 0 & 0 & 0 & \cdots \\
 i \lambda \theta_1 S^\dagger &0  & H_s - \mu_1 & 0 & 0 & \cdots \\
i \lambda \theta_2 S  & 0 & 0 & H_s  + \mu_2 &  0 & \cdots \\
i \lambda \theta_2 S^\dagger &0  & 0 & 0 & H_s - \mu_2& \cdots \\
\vdots & \vdots & \vdots  & \vdots & \vdots & \ddots
\end{array}\right].
\end{equation}
In addition, $H_1$ is skew-Hermitian.  $H_1$ and $V_k$ from \eqref{eq: ext0} are given by,
\begin{equation}\label{eq: Vk}
 H_1= -i \left(\begin{array}{cccccc} 0&  &  &  &  &    \\  & \nu_1 &  &  &  &    \\  &  & \nu_1 &  &  &    \\  &  &  & \nu_2 &  &    \\  &  &  &  & \nu_2 &    \\  &  &  &  & &  \ddots  \\ \end{array}\right), \quad  
 V_1 =\left(\begin{array}{c} 0  \\  0  \\    \sqrt{2 \nu_1}     \\  0      \\0     \\   \vdots     \\ \end{array}\right),  \quad 
 V_2 =\left(\begin{array}{c} 0  \\ 0      \\0     \\  0  \\    \sqrt{2 \nu_2}     \\     \vdots     \\ \end{array}\right), \cdots 
\end{equation}


The density-matrix for an SSE corresponds to the second-moment \cite{gisin1992quantum}.
For the extended system \eqref{eq: ext0}, we denote the density-matrix by $\Gamma$,  
\begin{equation}\Gamma=\left[
\begin{array}{cccc}
     \overline{    |\psi \rangle \langle \psi | } & \overline{ |\psi \rangle \langle \chi_1^\i |} & \overline{  |\psi \rangle \langle \chi_1^\ii | } & \cdots  \\
      \overline{   |\chi_1^\i \rangle \langle \psi |} &\overline{  |\chi_1^\i \rangle \langle \chi_1^\i |} & \overline{  |\chi_1^\i\rangle \langle \chi_1^\ii | }&  \cdots  \\
     \overline{    |\chi_1^\ii \rangle \langle \psi | }& \overline{  |\chi_1^\ii \rangle \langle \chi_1^\i | }& \overline{   |\chi_1^\ii \rangle \langle \chi_1^\ii |}&  \cdots  \\
        \vdots & \vdots & \vdots & \ddots 
           \end{array}\right]. 
\end{equation}

Therefore, the density-matrix associated with the combined wave functions $\Psi$ in \eqref{eq: Psi} follows the master equation,
\begin{equation}\label{eq: qme}
    \partial_t \Gamma = -i[\hat H_S, \Gamma]  +  \sum_{k=1}^{k_\text{max}}  V_k \rho V_k^\dagger.
\end{equation}
This can be verified by using the It\^o's formula [\onlinecite[Section 4.2]{kloeden2013numerical}]. 
In light of Eq. \eqref{eq: initchi}, the initial condition of the density-matrix can be written in a block form,
\begin{equation}\label{eq: gamo}
\Gamma(0)= \left(\begin{array}{cccccc}
\rho & 0 & 0 & 0 & 0 &\cdots\\
0 & 0 & 0 & 0 & 0 &\cdots\\
0 & 0 & \rho & 0 & 0 &\cdots\\
0 & 0 & 0 & 0 & 0 &\cdots\\
0 & 0 & 0 & 0 & \rho &\cdots \\
\vdots &\vdots &\vdots &\vdots &\vdots & \ddots
\end{array}\right)
\end{equation}

The QME that corresponds to the  type II embedding with correlated noise is slightly different.  Due to the fact that the auxiliary wave functions come in pairs and they share the same white noise,  we define,
\begin{equation}\label{eq: Vk'}
V_1 =\left(\begin{array}{c} 0  \\  0  \\    \sqrt{2 \nu_1}     \\  0      \\ \sqrt{2 \nu_2}     \\ 0  \\  0  \\0  \\  0  \\  \vdots     \\ \end{array}\right),  \quad 
 V_2 =\left(\begin{array}{c} 0  \\ 0      \\0     \\  0  \\    0    \\   0  \\  \sqrt{2 \nu_3} \\0  \\  \sqrt{2 \nu_4}  \\  \vdots     \\ \end{array}\right), \cdots,
 V_{k_\text{max}/2}=\left(\begin{array}{c} 0  \\ 0      \\0     \\  0  \\    0    \\   \vdots  \\  \sqrt{2 \nu_{k_\text{max}-1}} \\0  \\  \sqrt{2 \nu_{k_\text{max}}}  \\  \vdots     \\ \end{array}\right) 
\end{equation}

These operators  will enter the GQME \eqref{eq: qme}.  Due to the correlation, the initial density-matrix contains off-diagonal blocks,
 \begin{equation}\label{eq: qme2-init}
\Gamma(0)= \left(\begin{array}{cccccc}
\rho & 0 & 0 & 0 & 0 &\cdots\\
0 & 0 & 0 & 0 & 0 &\cdots\\
0 & 0 & \rho & 0 & Q_{12} \rho &\cdots\\
0 & 0 & 0 & 0 & 0 &\cdots\\
0 & 0 & Q_{12}^* \rho & 0 & \rho &\cdots \\
\vdots &\vdots &\vdots &\vdots &\vdots & \ddots
\end{array}\right).
\end{equation}

\subsection{Stochastic Propagator}
We briefly present a method for solving the extended SSEs \eqref{eq: ext0}. As emphasized in \cite{biele2012stochastic}, the construction of solution methods for stochastic models can be subtle. Due to the presence of the multiplicative noise in the SSEs \eqref{eq: sse-I'},  the conventional Euler-Maruyama method only has strong order 0.5 and weak order 1 \cite{kloeden2013numerical}. Here, one-step methods will be used, so it is enough to 
show how the equations are integrated from $t=0$ to $t=\Delta t$ with $\Delta t$ being the step size. 
A convenient approach to construct a numerical scheme is via operator-splitting \cite{li2020exponential,suzuki1993convergence,telatovich2017strong}. Let us combine the wave functions $\psi$ and $\chi_k$ into $\Psi.$ The idea is to  decompose the system \eqref{eq: sse-I'}
 into,
\[  i \partial_t \Psi = H \Psi,  \quad i \dot \zeta_k =0,  \]
for which the system becomes linear and the solution can be expressed using an exponential operator, 
\begin{equation}\label{eq: exp-int}
\Psi(\Delta t) = \exp (-i \Delta t H) \Psi(0),
\end{equation}
 and
\[ 
\partial_t \Psi=0, \quad  i \dot{\zeta}_k = -\alpha_k \zeta_k +  \gamma_k \dot{W}_k(t).  
\]

For the first part of the solution, one may use the Krylov subspace projection \cite{hochbruck1997krylov}, which has been shown to be very accurate and robust for solving time-dependent Schr\"{o}dinger equations \cite{castro2004propagators}. For the other part,
 a numerical solution can be constructed using the variation of constant formula \cite{ma2017fluctuation}, 
\begin{equation}
 \zeta_k(\Delta t)= \exp(i\alpha_k \Delta t) \zeta_k(0) +  \xi_k, \quad \xi_k=-i \sqrt{2\nu_k} \int_0^{\Delta t} \exp(i \alpha_k (t-\Delta t)) dW_k(t).
\end{equation}
The variable $\xi_k$ is a Gaussian random variable with zero mean. The variance can be obtained from the It\^o's isometry \cite{oksendal2003stochastic},
\[ \overline{\xi^2_k} = 2\nu_k \int_0^{\Delta t} \exp(2\nu_k (t-\Delta t) )dt=  1 - \exp( - 2\nu_k \Delta t).\]
Therefore the random variable $\xi_k$ can be sampled accordingly.

A more accurate splitting method can be obtained by solving the first part for half of the step, then the second part for one step, followed by another half step of the first problem \cite{yoshida1990construction}. This is known as a symmetric operator-splitting. Namely, we follow the flow chart,
\begin{equation}
\Psi \leftarrow \exp (-i \frac{\Delta t}2 H) \Psi \Rightarrow 
 \zeta_k(\Delta t)= \exp(i\alpha_k \Delta t) \zeta_k(0) +   \xi_k
 \Rightarrow  \Psi \leftarrow \exp (-i \frac{\Delta t}2 H) \Psi
 \end{equation}
The accuracy can be justified by expressing the solution of the SDEs using an exponential operator, followed by the Baker-Campbell-Hausdorff formula \cite{li2020exponential}.


\section{Example}

To test the embedded SSEs and QMEs for a non-Markovian dynamics, we follow the example by Gaspard and Nagaoka \cite{gaspard1999non} and consider a spin-boson model. In particular, we pick $H_s= -\frac{\Delta}2 \hat\sigma_z$, and the 
coupling operator $S=\hat\sigma_x$. The operator $B$ corresponds to the spectral density,
\begin{equation}
J(\omega)= \frac{\omega^3}{\omega_c^2}\exp \left(-\omega/\omega_c\right). 
\end{equation}

In this case, the correlation of the Gaussian noise is given by,
\begin{equation}\label{eq: C(t)}
c(t)= \int_0^{+\infty} \Big(\coth \frac{\beta \omega}{2} \cos \omega t - i \sin \omega t\Big). 
\end{equation}
Of particular importance to our approach is the power spectrum,  given by  \cite{gaspard1999non}, 
\begin{equation}\label{eq: gw2}
|G(\omega)|^2=\frac{\omega^3 e^{-|\omega|/\omega_c}}{\omega_c^2 (1- e^{-\beta \omega})}. 
\end{equation}
Notice that $|G(\omega)|^2 = \mathcal{O}(\omega^2)$ near zero.

We choose the model and numerical parameters according to the numerical experiment in \cite{gaspard1999non}, and they are  listed in Table \ref{para}. 
\begin{table}[thp]
\caption{\label{para} Parameters for the spin-boson model}
\begin{ruledtabular}
 \begin{tabular}{@{\hspace{3em}} ccccc@{\hspace{3em}} }
$ \Delta$ & $\omega_c$ & $\beta$ & $\lambda$ & $\Delta t$\\ 
 \hline
  0.1 & 1 & 10 & 0.1 & 0.05\\
 \end{tabular}
 \end{ruledtabular}
 \end{table}
 
 In the simulations, we monitor four quantities, including the $x$, $y$, $z$ components and the total mass $n:$
 \[ x = \text{tr}(\rho \hat\sigma_x), \quad y = \text{tr}(\rho \hat\sigma_y), \quad z = \text{tr}(\rho \hat\sigma_z), \quad n = \text{tr}(\rho).\]

\begin{figure}[tbp]
\begin{center}
\includegraphics[scale=0.18]{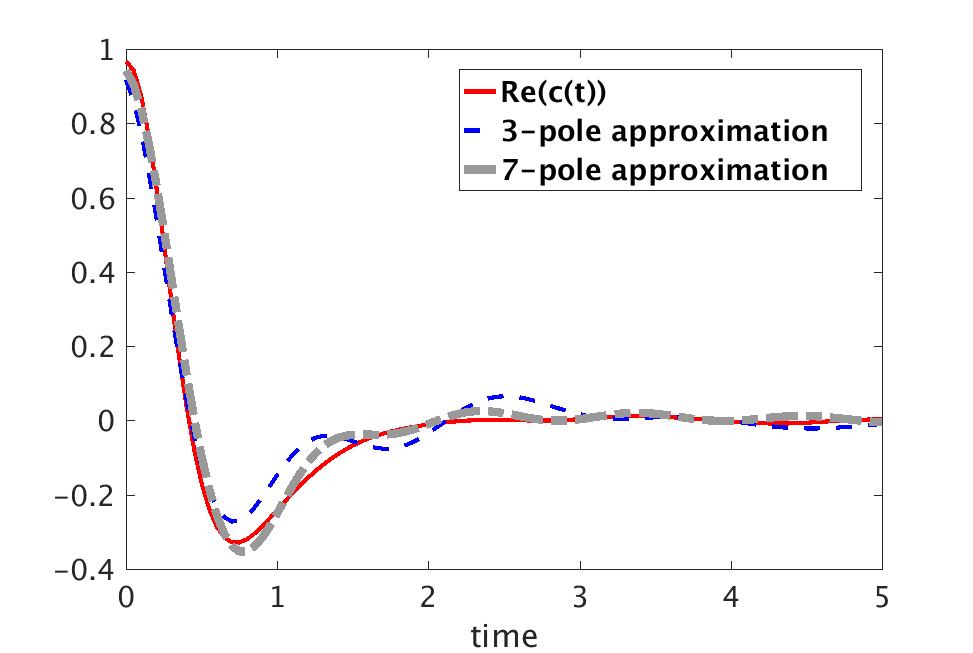}
\includegraphics[scale=0.18]{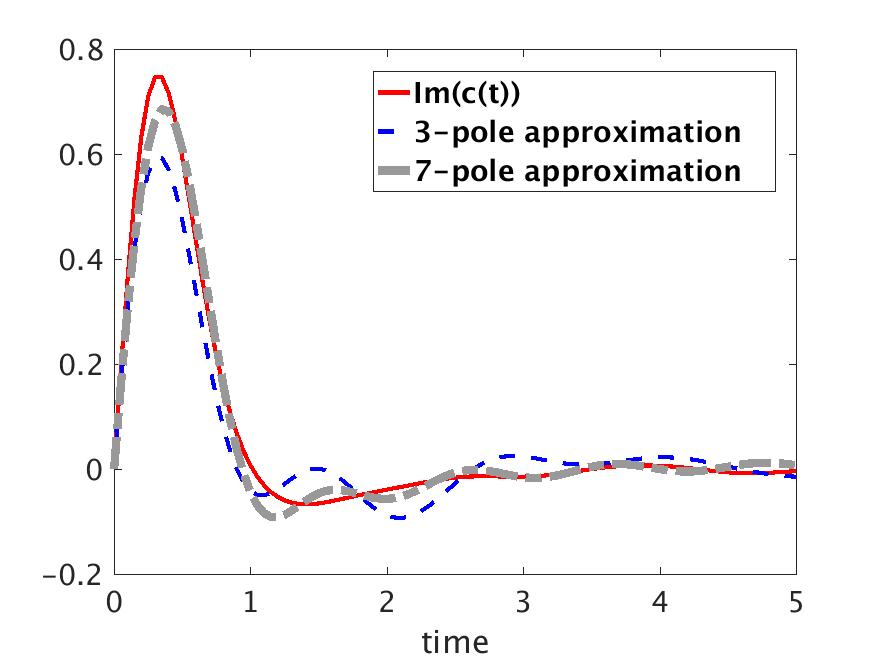}
\caption{The approximate time correlation function $c(t)$ from the ansatz  \eqref{eq: Gw}, compared to the exact correlation function. }
\label{fig: tcor}
\end{center}
\end{figure}

We first examine the approximation of the correlation function by \eqref{eq: sum-exp}. The parameters are computed using a nonlinear least-square algorithm, fitting the power spectrum \eqref{eq: gw2} by a sum of Lorentzians  \eqref{eq: Gw}.  The corresponding stochastic noise $\eta(t)$ is approximated by a sum of independent OU processes \eqref{eq: zeta2eta0}. Figure \ref{fig: tcor} shows the approximations using three poles ($k_{max}=3$) and seven poles ($k_{max}=7$).  The parameters are listed in table \ref{para1} for the three-pole approximation, and in table  \ref{para2} for the seven-pole approximation. 
  The exact correlation function $c(t)$ from Eq. \eqref{eq: C(t)} is computed from the power spectrum using fast Fourier transform (FFT), as explain in the Appendix. This calculation can also be done by using the Gauss-Laguerre quadrature due to the presence of the exponential term in $|G(\omega)|^2$.   With only three terms, the approximation  \eqref{eq: sum-exp} of the time correlation function $c(t)$ is already quite reasonable. The approximation using seven terms offers noticeable improvement. In this case, the nonlinear least-square fitting has multiple local minima. This issue is mitigated by re-running the nonlinear least-square algorithm with random initial perturbations. 

\begin{table}[thp]
\caption{\label{para1} Parameters from the 3-pole approximation}
\begin{ruledtabular}
 \begin{tabular}{@{\hspace{1em}}  c@{\hspace{1em}}   ccc @{\hspace{3em}} }
$ \theta_k $ &  0.4869  & 0.6986&   0.6407   \\
  \hline
 $\mu_k $ &  -2.0348  & -4.6671 &  -3.2467   \\
\hline
  $\nu_k$ & 0.4656 & 1.0061 & 0.6623 \\
 \end{tabular}
 \end{ruledtabular}
 \end{table}

\begin{table}[thp]
\caption{\label{para2} Parameters from the 7-pole approximation}
\begin{ruledtabular}
 \begin{tabular}{@{\hspace{1em}}  c@{\hspace{1em}}   ccccccc @{\hspace{2em}} }
$ \theta_k $ &0.2613 &      0.3353 &      0.3697&      0.3940 &      0.3911 &      0.3952 &       0.4005  \\
  \hline
 $\mu_k $ &  -1.4765  &    -2.0389 &     -2.5893 &      -3.1610 &      -3.7798 &     -4.4889 &     -5.4372  \\
\hline
  $\nu_k$ &  0.3253  &    0.3599 &      0.3881 &        0.4320 &      0.4716 &      0.5406 &      0.6207 \\
 \end{tabular}
 \end{ruledtabular}
 \end{table}

To examine the embedded SSEs, we first solve the non-Markovian SSE \eqref{eq: sse} using a scheme suggested in \cite{gaspard1999non}.  More specifically, the memory term is approximated by a composite trapezoid rule. More specifically, we first use the fast Fourier transform (FFT) to compute the correlation function $c(t)$ from the power spectrum \eqref{eq: gw2}. The noise can be computed by sampling independent Gaussian variables according to the power spectrum, followed by an inverse FFT, as explained in Appendix C.

\begin{figure}[tph]
\begin{center}
\includegraphics[scale=0.18]{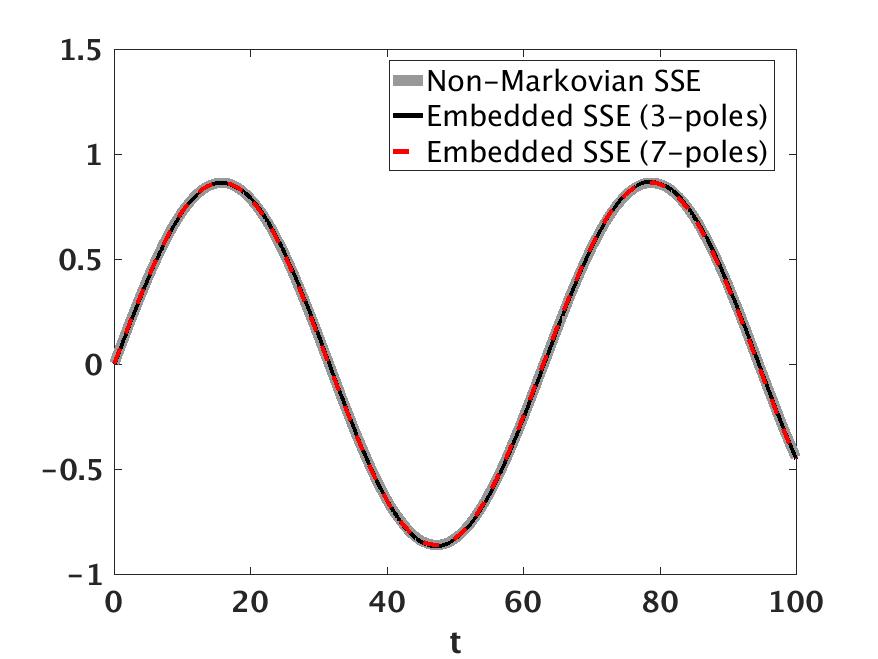}
\includegraphics[scale=0.18]{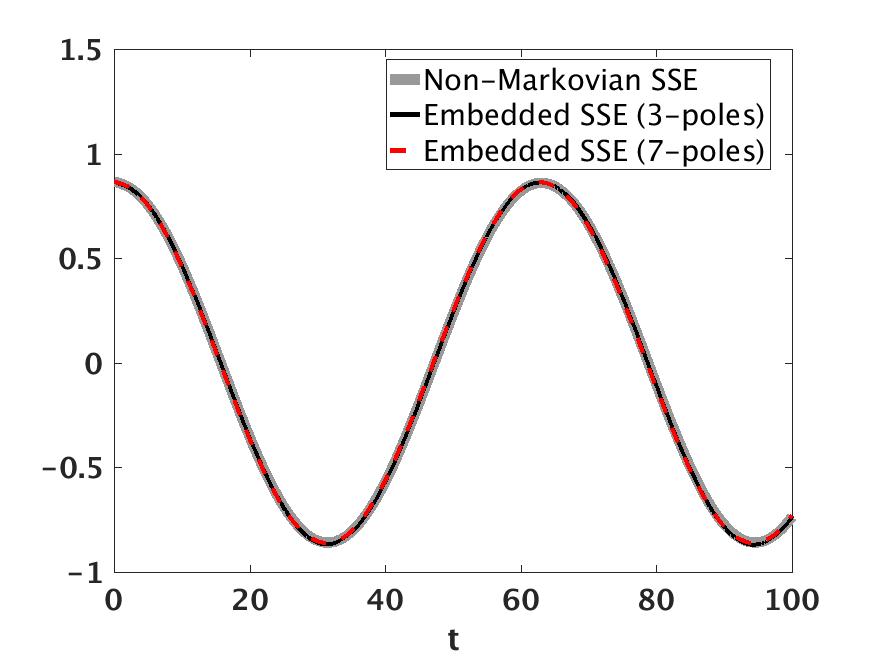}\\
\includegraphics[scale=0.18]{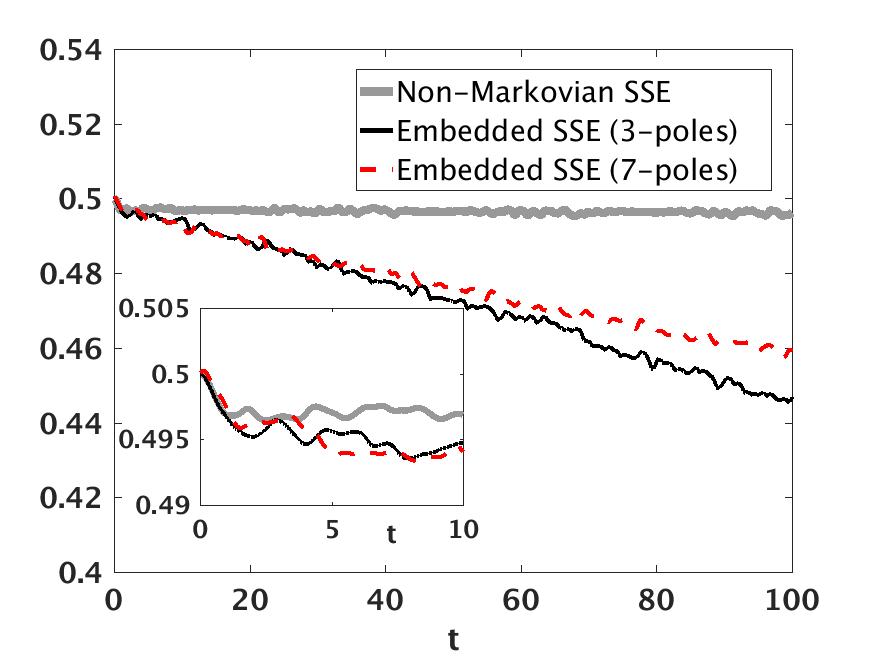}
\includegraphics[scale=0.18]{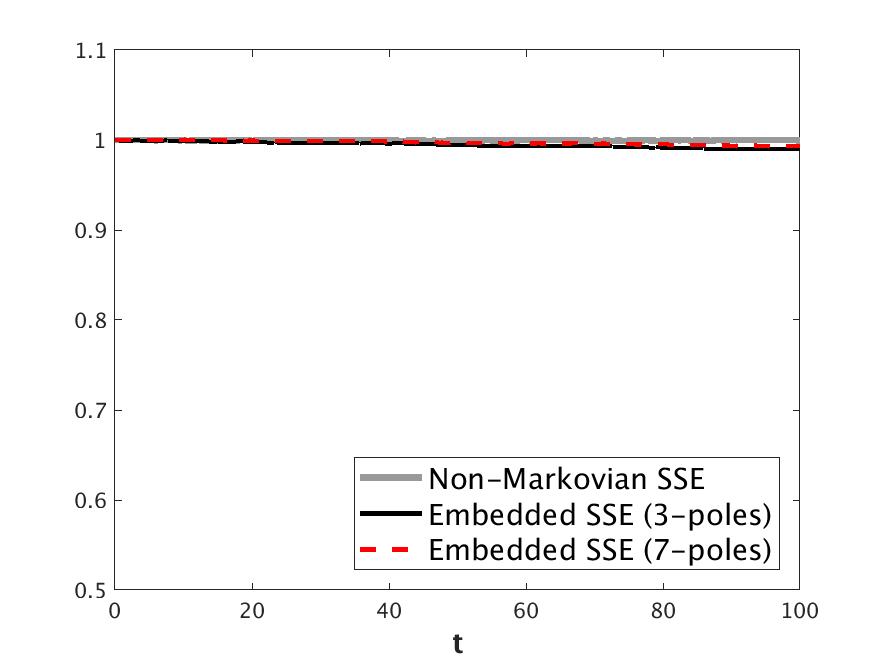}
\caption{Numerical results from the extended SSE  \eqref{eq: sse-I}, compared to those from the original non-Markovian SSE \eqref{eq: sse}. The averages are computed using $10^4$ independent copies. }
\label{fig: avg}
\end{center}
\end{figure}

We now solve the type {\bf I} extended SSEs \eqref{eq: sse-I}.  Figure \ref{fig: avg} shows the values of $x$, $y$ and $z$, along with the total mass $n$ in time, computed from both models.  We see that the results agree very well except for the dynamics of $z$. We will return to this point later. The total mass predicted using seven poles show slight improvement.

\begin{figure}[htp]
\begin{center}
\includegraphics[scale=0.18]{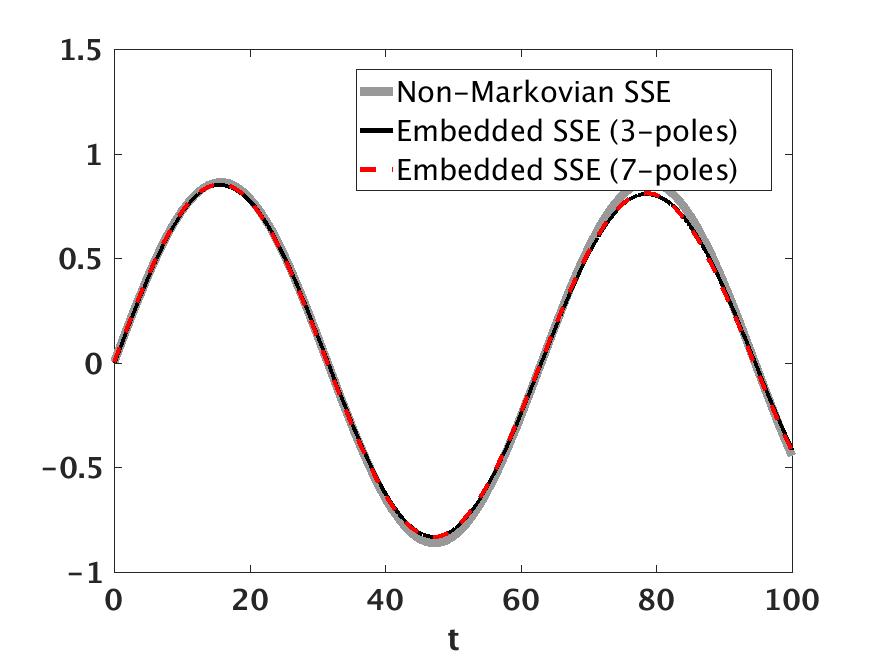}
\includegraphics[scale=0.18]{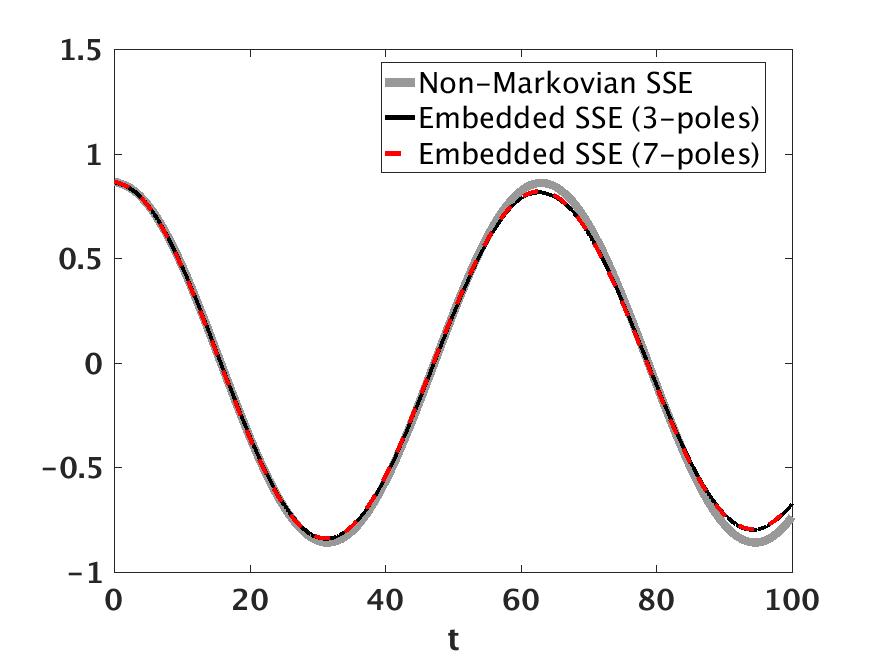}\\
\includegraphics[scale=0.18]{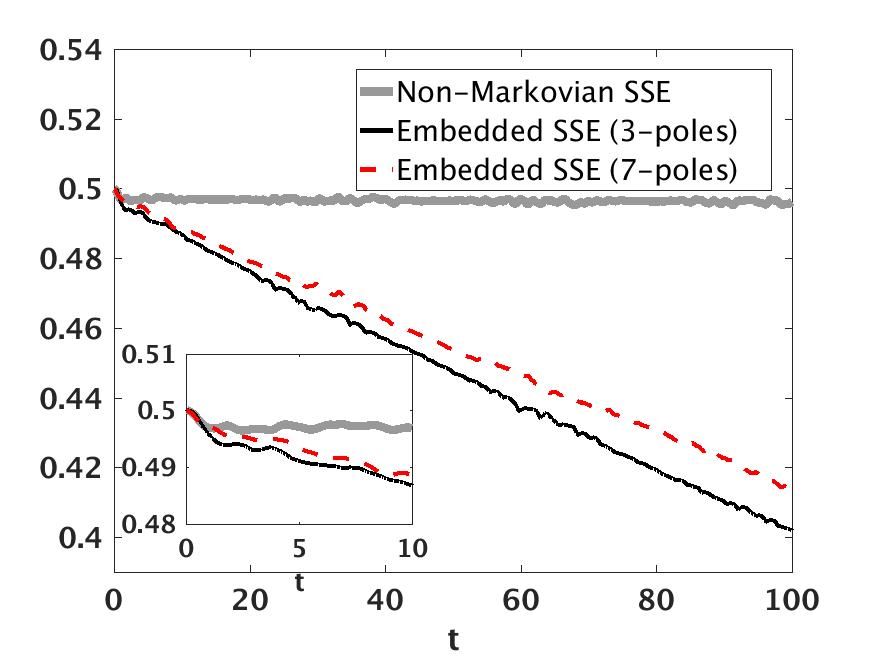}
\includegraphics[scale=0.18]{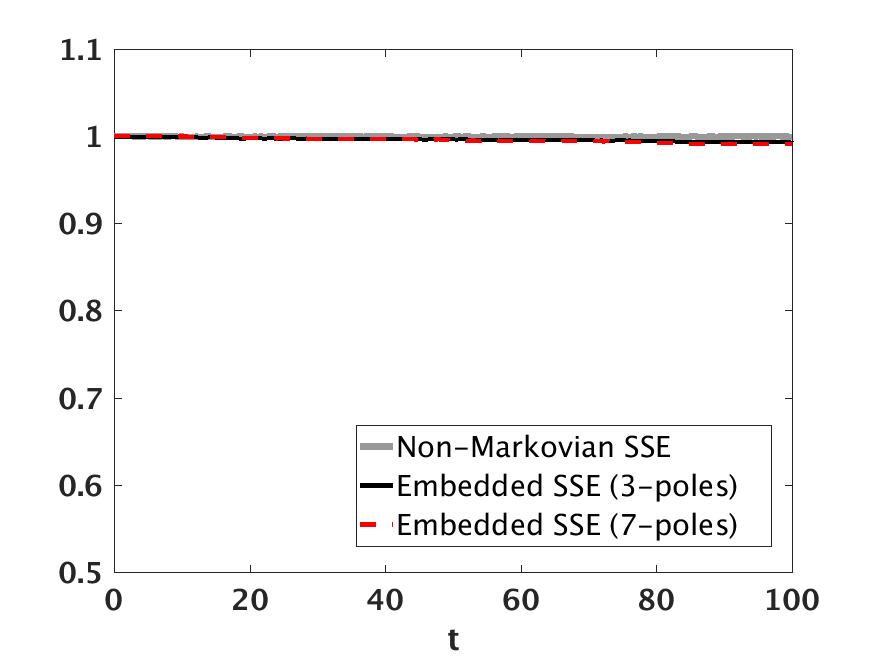}
\caption{Numerical results from the QME  \eqref{eq: qme}, compared to those from the original non-Markovian SSE \eqref{eq: sse}. }
\label{fig: avg1}
\end{center}
\end{figure}

For the type {\bf II} embedding, we choose to solve the corresponding GQME  \eqref{eq: qme}, which in this case, is much more efficient than
embedded SSEs \eqref{eq: ext0}, since we do not have to run many realizations.  We follow the operator \eqref{eq: Vk} and the initial condition  \eqref{eq: gamo}.  Figure \ref{fig: avg1} shows the predicted values of $x$, $y$, $z$ and the mass. Again the results agree well with those obtained from the original non-Markovian SSE \eqref{eq: sse}, except for the $z$ component. 

\bigskip 

\begin{table}[thp]
\caption{\label{para3} Parameters from the four-term approximation}
\begin{ruledtabular}
 \begin{tabular}{@{\hspace{1em}}  c@{\hspace{1em}}   cccccccc @{\hspace{2em}} }
$ Q_k $ & 0.0485  &  0.0175 &   0.1118 &   0.3016 & & & & \\
  \hline
 $\mu_k $ &  -1.3285&   -2.1347&   -0.4553&   -1.1296&   -3.4772&   -2.3696&   -5.8125&   -3.8587 \\
\hline
  $\nu_k$ & 0.4638  &   0.5939 &   0.3338 &   0.3776  &  0.9486 &   0.6595 &   1.9818 &   1.0053 \\
 \end{tabular}
 \end{ruledtabular}
 \end{table}

 The poor predictions of $z(t)$  by the previous approximations can be attributed to the inconsistency with the power spectrum near $\omega=0.$ Since the true spectrum scales like $\omega^2$ near zero, an appropriate approximation should take this into account. Preserving such asymptotics has been emphasized in many works \cite{meier_non-markovian_1999,ritschel2014analytic,sus_hierarchy_2014}.   Here we follow the ansatz \eqref{eq: Gwc}, which at the level of stochastic processes, can be realized by a linear combination of the correlated OU processes \eqref{eq: ck2eta}. For this example, we use the ansatz \eqref{eq: Gwc} with four terms ($k_{max}=8$). The values in \eqref{eq: Gwc} obtained from the fitting procedure are listed in Table \ref{para3}.

\begin{figure}[htp]
\begin{center}
\includegraphics[scale=0.22]{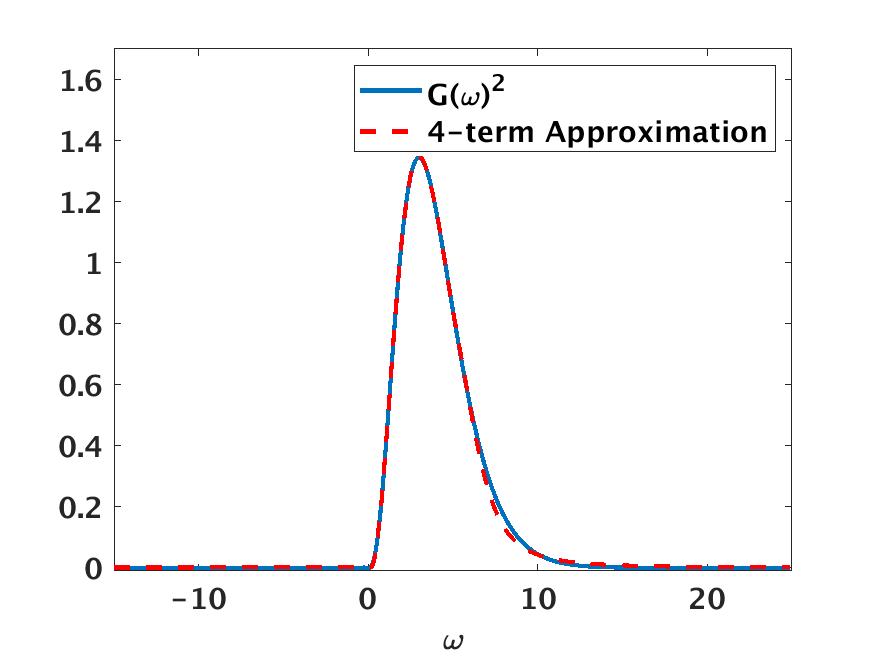}
\caption{The approximation of the power spectrum  \eqref{eq: Gw} using the ansatz  \eqref{eq: Gwc} with four terms. }
\label{fig: gw}
\end{center}
\end{figure}

\begin{figure}[htp]
\begin{center}
\includegraphics[scale=0.18]{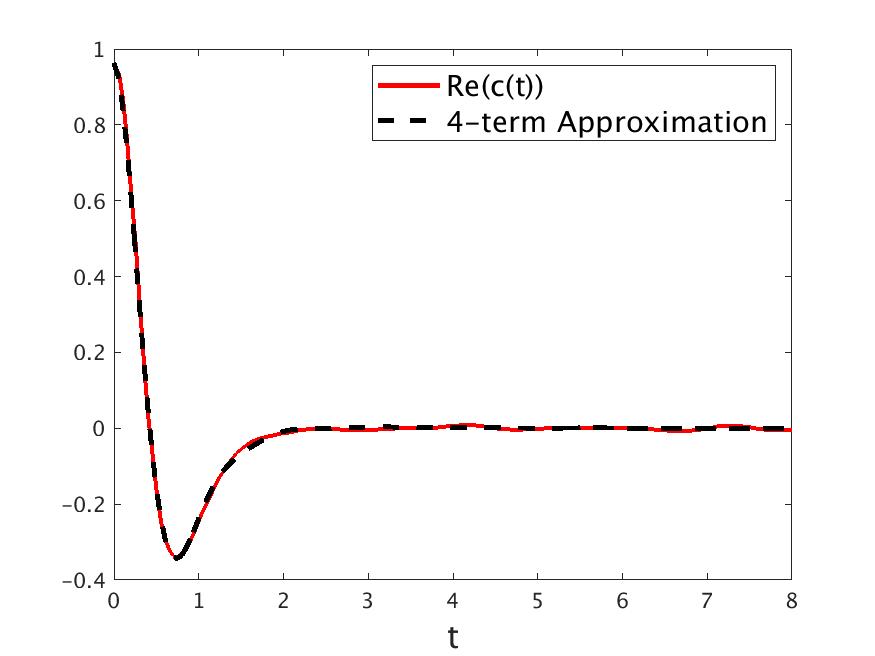}
\includegraphics[scale=0.18]{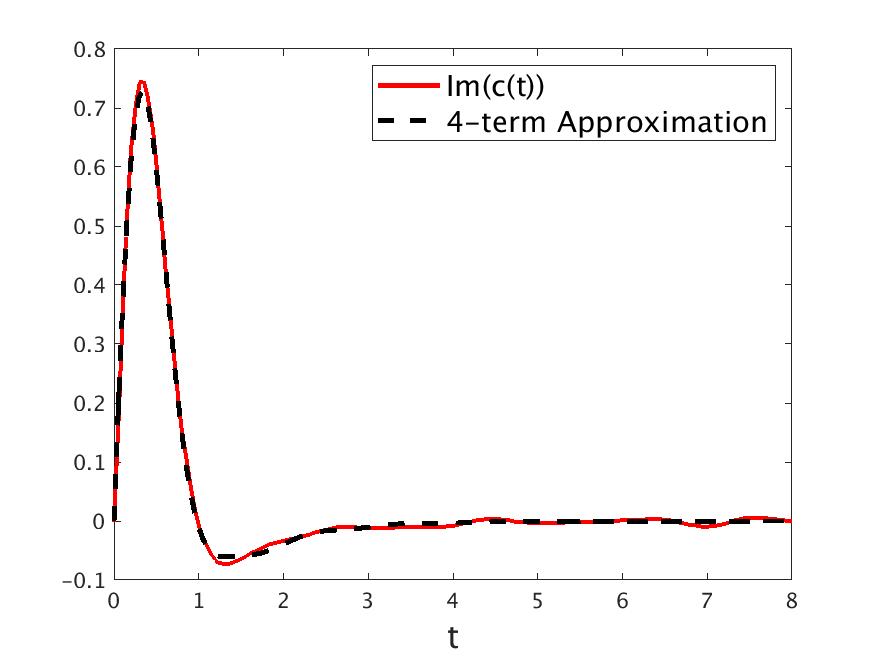}
\caption{The approximation of the time correlation function $C(t)$ using the ansatz  \eqref{eq: Gwc} with four terms. The approximate time correlation function is obtained from {eq: torc-all}.}
\label{fig: tcorc}
\end{center}
\end{figure}

 Figure \ref{fig: gw} displays the approximation of the power spectrum, which clearly shows consistency at $\omega=0.$ This improved accuracy is also reflected in the time correlation function, as shown in Figure \ref{fig: tcorc}. The approximation of the correlation function is  significantly improved compared to the ansatz of sum of Lorentzians  \eqref{eq: Gw} (Figure \ref{fig: tcor}).  

\begin{figure}[htp]
\begin{center}
\includegraphics[scale=0.18]{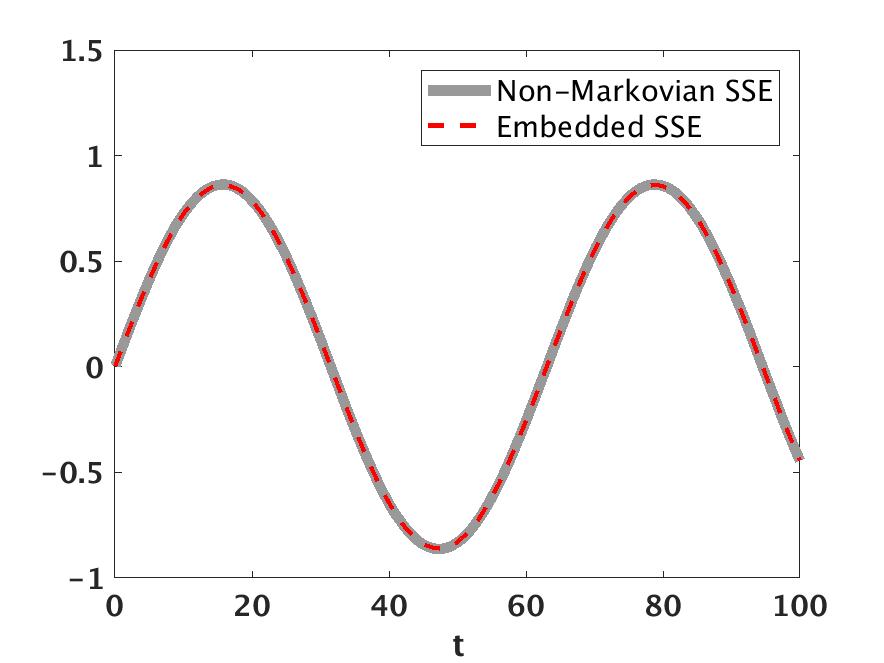}
\includegraphics[scale=0.18]{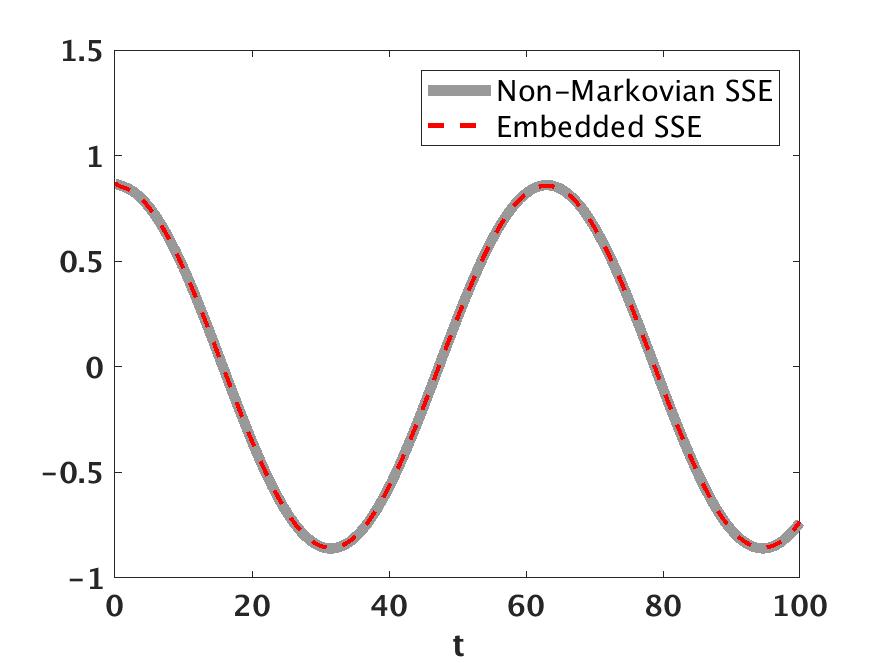}\\
\includegraphics[scale=0.18]{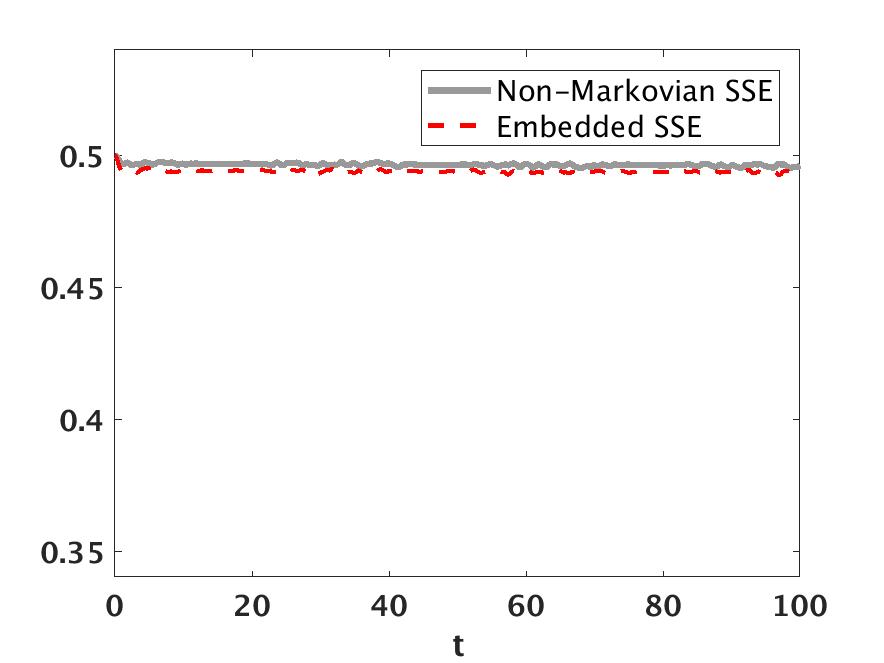}
\includegraphics[scale=0.18]{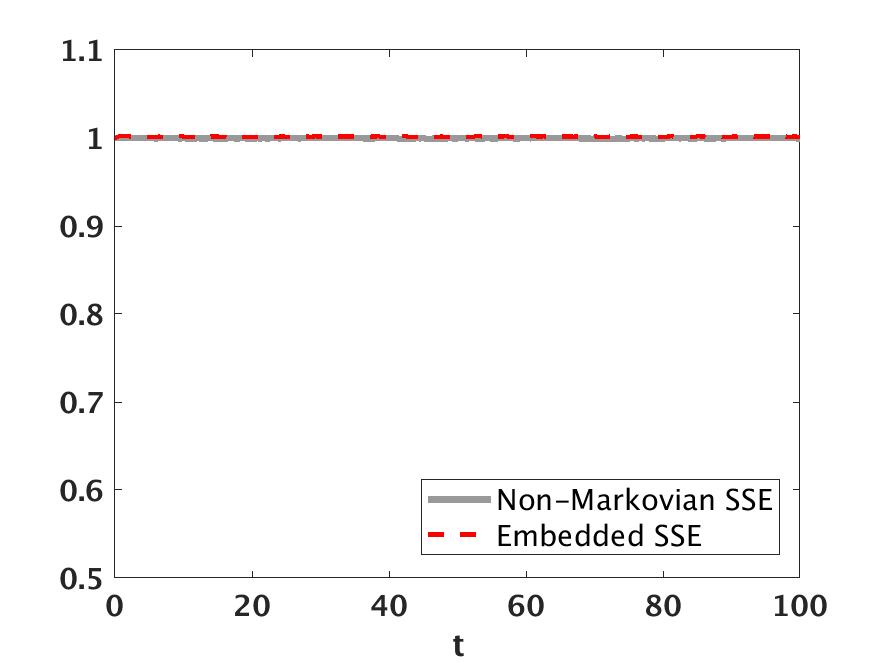}
\caption{Numerical results from the extended SSE  \eqref{eq: sse-I''}, compared to those from the original non-Markovian SSE \eqref{eq: sse}. The averages are computed using $10^4$ independent copies. }
\label{fig: avgc1}
\end{center}
\end{figure}

With this four-term approximation, we solved the type {\bf I} embedded SSEs  \eqref{eq: sse-I''}. The results, as depicted in Figure \ref{fig: avgc1}, exhibit great accuracy in the prediction of all four quantities. 

To test this four-term approximation on the type {\bf II} embedded SSEs \eqref{eq: sse-II''}, we solve the corresponding GQME \eqref{eq: qme}, together with the operators $V_k$ given by \eqref{eq: Vk'} and initial condition \eqref{eq: qme2-init}. As shown in Figure \ref{fig: avgc1}, all the four quantities are well captured as well. Compared to solving the SSEs  \eqref{eq: sse-I''} and  \eqref{eq: sse-II''}, it takes a lot less  CPU time to solve the GQME, since no ensemble average is needed. 
\begin{figure}[htp]
\begin{center}
\includegraphics[scale=0.18]{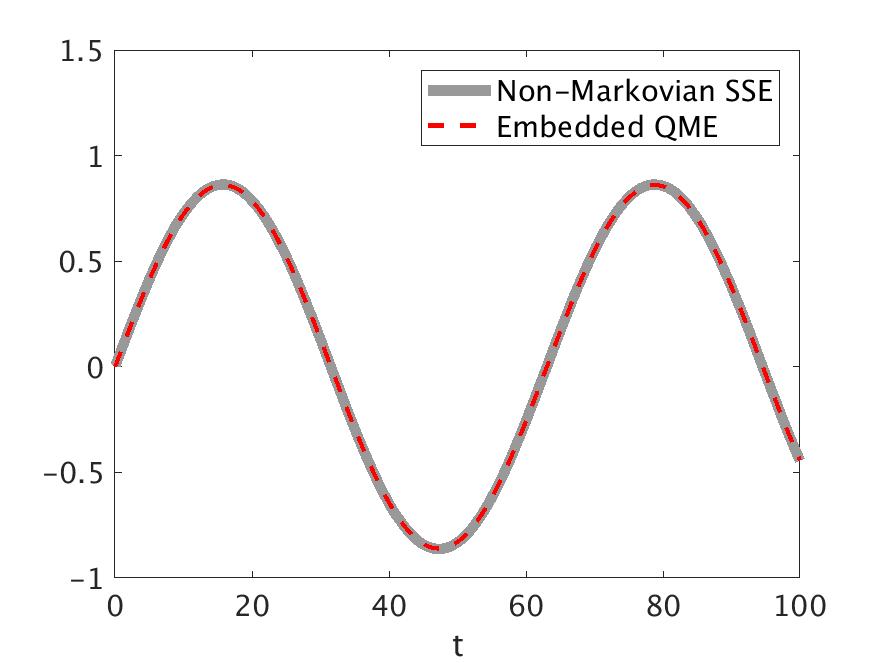}
\includegraphics[scale=0.18]{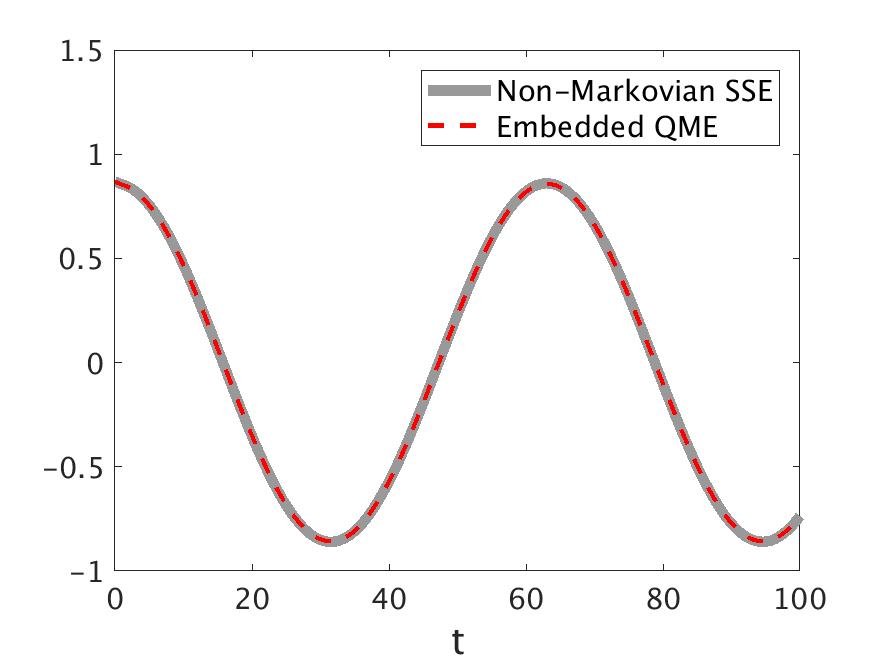}\\
\includegraphics[scale=0.18]{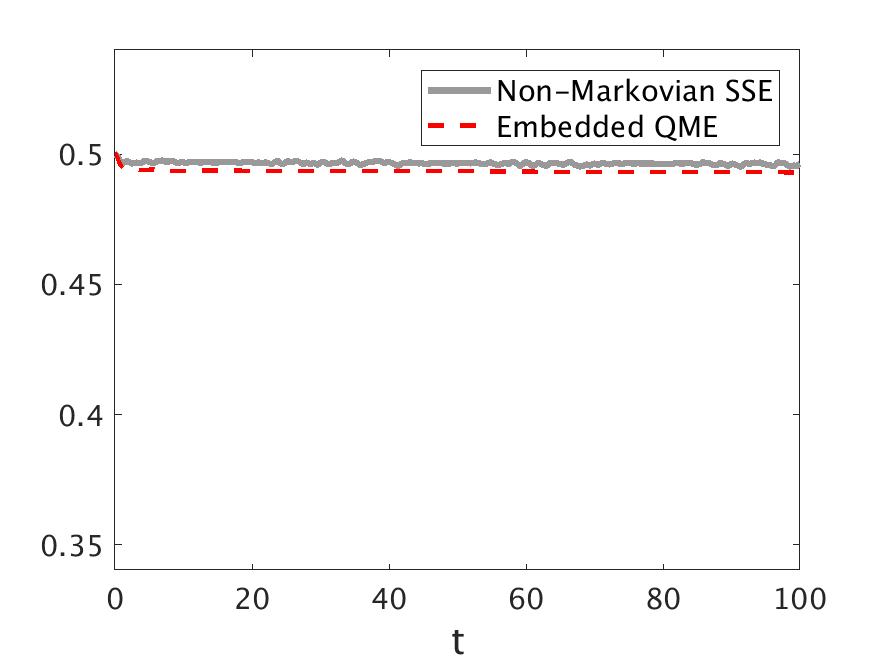}
\includegraphics[scale=0.18]{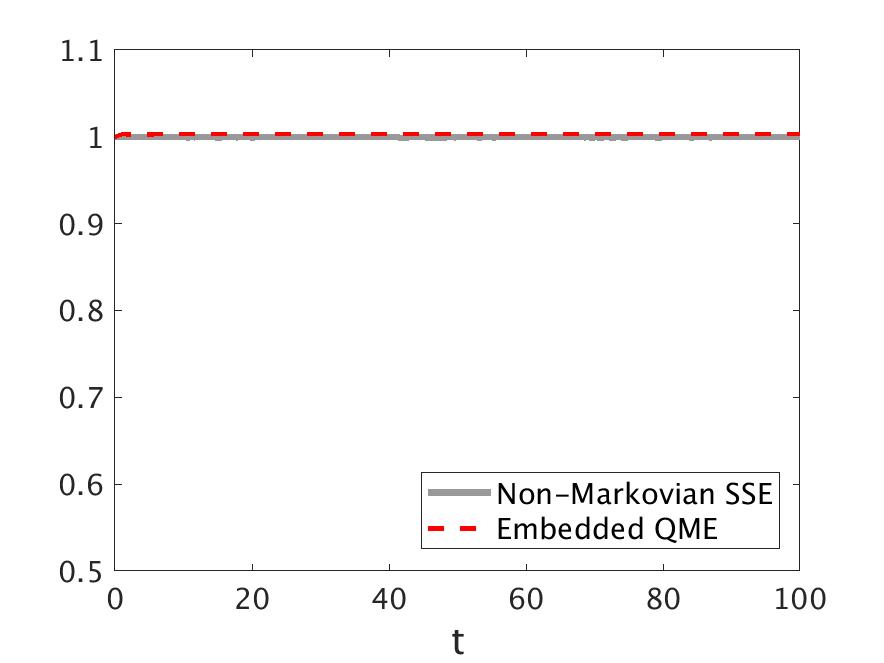}
\caption{Numerical results from the QME QME  \eqref{eq: qme}, derived from the type {\bf II} embedding  \eqref{eq: ext0}, compared to those from the original non-Markovian SSE \eqref{eq: sse}. The averages from \eqref{eq: sse} are computed using $10^4$ independent copies. }
\label{fig: avgc2}
\end{center}
\end{figure}

\section{Summary and Discussions}
We introduced two techniques to embed non-Markovian stochastic Schr\"odinger equations in an extended stochastic dynamics that can be characterized more precisely by It\^o calculus \cite{oksendal2003stochastic}. The second type of embedding also yields generalized quantum master equations for the density-matrix.
The two descriptions using wave functions and density-matrix are constructed to be  complementary alternative with practical implementations in mind.  

The embedded SSEs are stochastic differential equations with explicit model parameters. They come from the properties of the bath, as well as the coupling with the system. One important application would be the parameter estimation problem \cite{banchi2018modelling}: Given the time series of physical observables, can one determine the model parameters with good statistical certainty?  This has been done for classical systems modeled by the generalized Langevin equation  \cite{chorin2015discrete,lei2016generalized}. Since the SSEs are now written as  Markovian stochastic differential equations, a variety of methods are available for this purpose \cite{bishwal2007parameter}. One suitable method is the Kalman filter for systems with multiplicative noise \cite{song2016linear}.

\begin{acknowledgments}
The author would like to acknowledge the support of National Science Foundation Grant DMS-1819011.
\end{acknowledgments}

\appendix

\section{The complex-valued OU process}
By separating the real and imaginary parts we can first turn the complex OU process  \eqref{eq: cou} into a real-valued two-dimensional SDE,
\begin{equation}
d \left[
    \begin{array}{c}
         \text{Re} (\zeta)  \\
         \text{Im}(\zeta)
    \end{array}\right] 
    = 
    \left[
    \begin{array}{cc}
         - \nu & -\mu \\
         \mu & - \nu
    \end{array}\right] 
     \left[
    \begin{array}{c}
         \text{Re} (\zeta)  \\
         \text{Im}(\zeta)
    \end{array}\right]  dt
    + \gamma \frac{1}{\sqrt{2}} 
    \left[
    \begin{array}{c}
         dW_1  \\
         dW_2 
    \end{array}\right]. 
\end{equation}
Here $\alpha = \mu + i \nu$, and $W(t)= \frac{1}{\sqrt{2}} (W_1 + i W_2).$ $W_1(t)$ and $W_2(t)$ are real-valued independent Brownian motions.

At this point, we can use the Lyapunov equation to determine the diffusion term [\onlinecite[Section 3.7~]{Pav_book:14}]. If we set,
\[Q=
\overline{
 \left[
    \begin{array}{c}
         \text{Re} (\zeta)  \\
         \text{Im}(\zeta)
    \end{array}\right] 
    \left[
    \begin{array}{cc}
      \text{Re} (\zeta)  &
         \text{Im}(\zeta)
    \end{array}\right]
    }, 
\]
then the Lyapunov equation gives,
\[ 
  \left[
    \begin{array}{cc}
         - \nu & -\mu \\
         \mu & - \nu
    \end{array}\right] Q + Q
      \left[
    \begin{array}{cc}
         - \nu & -\mu \\
         \mu & - \nu
    \end{array}\right] + 
 \left[
    \begin{array}{cc}
         \frac{\gamma^2}{2} &  \\
          &\frac{\gamma^2}{2}
    \end{array}\right] =0.\]
This gives the relation $Q=I$ and
\[ \gamma^2 = 2\nu. \]

\section{Correlated OU processes}

We can build a new Gaussian process using two correlated OU processes as follows,
\begin{equation}\label{eq: eta12}
\left\{
\begin{aligned}
  i \partial_t \zeta_1 =& - \alpha_1 \zeta_1 + \gamma_1 \dot{W}(t), \\
  i \partial_t \zeta_2 =& - \alpha_2 \zeta_2 + \gamma_2 \dot{W}(t). \\ 
\end{aligned}\right.
\end{equation} 
Without loss of generality, we consider the construction of $L_1(\omega)$ in \eqref{eq: Gwc}.
We first make the normalization, as suggested in the previous section, that $\gamma_i^2 = 2 \text{Im}(\alpha_i), \;i=1, 2.$ It is important to note that to enable the correlation, the two processes are driven by the same white noise $\dot{W}(t)$. 

Using the It\^{o} formula, we have for the covariance matrix $Q$, with $Q_{i,j}= \overline{\zeta_i(t)^* \zeta_j(t)}$,
\begin{equation}
\frac{d}{dt} Q = Q 
\left(\begin{array}{cc}i \alpha_1 & 0 \\0 & i\alpha_2\end{array}\right)
+
\left(\begin{array}{cc} -i\alpha_1^* & 0 \\0 & -i \alpha_2^* \end{array}\right) Q +
\left(\begin{array}{cc} \gamma_1^2 & \gamma_1\gamma_2 \\
\gamma_1\gamma_2 & \gamma_2^2 \end{array}\right).
\end{equation}

Direct calculations yield the equilibrium covariance,
\begin{equation}
Q_{11}=Q_{22}=1, \quad  Q_{12}= \frac{\gamma_1 \gamma_2}{\alpha_2 -\alpha_1^*}, \quad  Q_{21}=Q_{12}^*.
\end{equation}

One can use the Fourier transform to solve the  SDEs \eqref{eq: eta12} [\onlinecite[Section 3.2]{risken1984fokker}],
\[ \hat \zeta_i(\omega) = \frac{\gamma_i}{\omega+ \alpha_i}  \hat{W}(\omega), \; i=1, 2. \]
With a linear combination, $\zeta= c_1 \zeta_1 + c_2 \zeta_2, $ we have,
\[
\hat{\zeta}(\omega)= \frac{ (c_1 \gamma_1 + c_2 \gamma_2) \omega + c_1 \gamma_1 \alpha_2 + c_2\gamma_2 \alpha_1}{( \omega+ \alpha_1)(\omega + \alpha_2)}, 
\]
which becomes,
\begin{equation}
\hat{\zeta}= \frac{ \omega }{( \omega+ \alpha_1)(\omega + \alpha_2)},
\end{equation}
if the coefficients are selected according to \eqref{eq: c12}. As a result, the power spectrum of $\zeta$ agrees with $L_1(\omega),$
\begin{equation}
 \overline{ \hat\zeta^* (\omega)\hat \zeta(\omega') } = L_1(\omega) \delta(\omega - \omega').
\end{equation}

\section{The FFT approach to sample the noise $\eta(t)$ from $c(t)$}

We first consider the inverse Fourier transform,
\begin{equation}\label{eq: if}
    f(t)= \mathcal{F}^{-1}\Big[F\Big]:=\frac{1}{2\pi} \int_{-\infty}^{+\infty} F(\omega) e^{i\omega t} d\omega. 
\end{equation}

If $F(\omega)=|G(\omega)|^2$, then the corresponding inverse Fourier transform corresponds to the correlation function $c(t).$ Namely,
\begin{equation}
    c(t) = \mathcal{F}^{-1}\Big[ |G|^2\Big].
\end{equation}
Now let $\xi(\omega)$ be the complex-valued white noise, i.e., 
$\overline{\xi(\omega)^*\xi(\omega')}= \delta (\omega - \omega').$
We may define a Gaussian process in the Fourier space by $G(\omega)\xi(\omega)$, followed by the inverse Fourier transform, 
\begin{equation}\label{eq: fft-eta}
    \eta(t) = \sqrt{2\pi} \mathcal{F}^{-1} \Big[ G(\omega)\xi(\omega) \Big]. 
\end{equation}
With direct calculations, one finds that,
\[ \overline{\eta(t)^* \eta(t')} = c(t-t'). \] 

\medskip

To make use of this relation using the fast Fourier transform (FFT), we pick $N$ equally spaced frequency $\omega_k= -\omega_{max} + 2 \pi(k-1)/(N \Delta t),$ with $\Delta t = \pi/\omega_{max}.$  We also define $\omega \omega= 2 \omega_{max}/N.$ 

Given the values $\{F(\omega_k)\}$,  $f(j\Delta t) \approx f_j $ is given by,
\begin{equation}\label{eq: IFFT}
    f_j = \frac{e^{-i\omega_{max} j \Delta t} }{\Delta t} \times \text{IFFT}(F), \quad \text{IFFT}(F)= \frac{1}{N}
     \sum_{k=1}^N F(\omega_k) e^{i2\pi (j-1)(k-1)/N}.
 \end{equation}
 This corresponds to a quadrature for the Fourier integral \eqref{eq: if}. 

 To implement \eqref{eq: fft-eta}, we pick the Gaussian processes $\xi(\omega_k)$ such that, 
 $$\overline{\xi(\omega_k)^*\xi(\omega_\ell)}=  \delta_{k\ell}/\Delta \omega.$$
 
 Then the noise at discrete time steps, in light of \eqref{eq: fft-eta} and \eqref{eq: IFFT},  can be generated as,
 \begin{equation}
    \eta(t_j) =\sqrt{2\pi}  \frac{e^{-i\omega_{max} j \Delta t} }{\Delta t} \times \text{IFFT}(G\xi).
 \end{equation}

\bibliography{nmsse}

\begin{thebibliography}{56}%
\makeatletter
\providecommand \@ifxundefined [1]{%
 \@ifx{#1\undefined}
}%
\providecommand \@ifnum [1]{%
 \ifnum #1\expandafter \@firstoftwo
 \else \expandafter \@secondoftwo
 \fi
}%
\providecommand \@ifx [1]{%
 \ifx #1\expandafter \@firstoftwo
 \else \expandafter \@secondoftwo
 \fi
}%
\providecommand \natexlab [1]{#1}%
\providecommand \enquote  [1]{``#1''}%
\providecommand \bibnamefont  [1]{#1}%
\providecommand \bibfnamefont [1]{#1}%
\providecommand \citenamefont [1]{#1}%
\providecommand \href@noop [0]{\@secondoftwo}%
\providecommand \href [0]{\begingroup \@sanitize@url \@href}%
\providecommand \@href[1]{\@@startlink{#1}\@@href}%
\providecommand \@@href[1]{\endgroup#1\@@endlink}%
\providecommand \@sanitize@url [0]{\catcode `\\12\catcode `\$12\catcode
  `\&12\catcode `\#12\catcode `\^12\catcode `\_12\catcode `\%12\relax}%
\providecommand \@@startlink[1]{}%
\providecommand \@@endlink[0]{}%
\providecommand \url  [0]{\begingroup\@sanitize@url \@url }%
\providecommand \@url [1]{\endgroup\@href {#1}{\urlprefix }}%
\providecommand \urlprefix  [0]{URL }%
\providecommand \Eprint [0]{\href }%
\providecommand \doibase [0]{http://dx.doi.org/}%
\providecommand \selectlanguage [0]{\@gobble}%
\providecommand \bibinfo  [0]{\@secondoftwo}%
\providecommand \bibfield  [0]{\@secondoftwo}%
\providecommand \translation [1]{[#1]}%
\providecommand \BibitemOpen [0]{}%
\providecommand \bibitemStop [0]{}%
\providecommand \bibitemNoStop [0]{.\EOS\space}%
\providecommand \EOS [0]{\spacefactor3000\relax}%
\providecommand \BibitemShut  [1]{\csname bibitem#1\endcsname}%
\let\auto@bib@innerbib\@empty
\bibitem [{\citenamefont {Gr{\"o}blacher}\ \emph {et~al.}(2015)\citenamefont
  {Gr{\"o}blacher}, \citenamefont {Trubarov}, \citenamefont {Prigge},
  \citenamefont {Cole}, \citenamefont {Aspelmeyer},\ and\ \citenamefont
  {Eisert}}]{groblacher2015observation}%
  \BibitemOpen
  \bibfield  {author} {\bibinfo {author} {\bibfnamefont {S.}~\bibnamefont
  {Gr{\"o}blacher}}, \bibinfo {author} {\bibfnamefont {A.}~\bibnamefont
  {Trubarov}}, \bibinfo {author} {\bibfnamefont {N.}~\bibnamefont {Prigge}},
  \bibinfo {author} {\bibfnamefont {G.}~\bibnamefont {Cole}}, \bibinfo {author}
  {\bibfnamefont {M.}~\bibnamefont {Aspelmeyer}}, \ and\ \bibinfo {author}
  {\bibfnamefont {J.}~\bibnamefont {Eisert}},\ }\bibfield  {title} {\enquote
  {\bibinfo {title} {Observation of {non-Markovian} micromechanical {Brownian}
  motion},}\ }\href@noop {} {\bibfield  {journal} {\bibinfo  {journal} {Nature
  Communications}\ }\textbf {\bibinfo {volume} {6}},\ \bibinfo {pages} {7606}
  (\bibinfo {year} {2015})}\BibitemShut {NoStop}%
\bibitem [{\citenamefont {Madsen}\ \emph {et~al.}(2011)\citenamefont {Madsen},
  \citenamefont {Ates}, \citenamefont {Lund-Hansen}, \citenamefont
  {L{\"o}ffler}, \citenamefont {Reitzenstein}, \citenamefont {Forchel},\ and\
  \citenamefont {Lodahl}}]{madsen2011observation}%
  \BibitemOpen
  \bibfield  {author} {\bibinfo {author} {\bibfnamefont {K.~H.}\ \bibnamefont
  {Madsen}}, \bibinfo {author} {\bibfnamefont {S.}~\bibnamefont {Ates}},
  \bibinfo {author} {\bibfnamefont {T.}~\bibnamefont {Lund-Hansen}}, \bibinfo
  {author} {\bibfnamefont {A.}~\bibnamefont {L{\"o}ffler}}, \bibinfo {author}
  {\bibfnamefont {S.}~\bibnamefont {Reitzenstein}}, \bibinfo {author}
  {\bibfnamefont {A.}~\bibnamefont {Forchel}}, \ and\ \bibinfo {author}
  {\bibfnamefont {P.}~\bibnamefont {Lodahl}},\ }\bibfield  {title} {\enquote
  {\bibinfo {title} {Observation of {non-Markovian} dynamics of a single
  quantum dot in a micropillar cavity},}\ }\href@noop {} {\bibfield  {journal}
  {\bibinfo  {journal} {Physical Review Letters}\ }\textbf {\bibinfo {volume}
  {106}},\ \bibinfo {pages} {233601} (\bibinfo {year} {2011})}\BibitemShut
  {NoStop}%
\bibitem [{\citenamefont {Breuer}\ \emph {et~al.}(2016)\citenamefont {Breuer},
  \citenamefont {Laine}, \citenamefont {Piilo},\ and\ \citenamefont
  {Vacchini}}]{breuer_non-markovian_2016}%
  \BibitemOpen
  \bibfield  {author} {\bibinfo {author} {\bibfnamefont {H.-P.}\ \bibnamefont
  {Breuer}}, \bibinfo {author} {\bibfnamefont {E.-M.}\ \bibnamefont {Laine}},
  \bibinfo {author} {\bibfnamefont {J.}~\bibnamefont {Piilo}}, \ and\ \bibinfo
  {author} {\bibfnamefont {B.}~\bibnamefont {Vacchini}},\ }\bibfield  {title}
  {\enquote {\bibinfo {title} {Non-{Markovian} dynamics in open quantum
  systems},}\ }\href@noop {} {\bibfield  {journal} {\bibinfo  {journal}
  {Reviews of Modern Physics}\ }\textbf {\bibinfo {volume} {88}} (\bibinfo
  {year} {2016})}\BibitemShut {NoStop}%
\bibitem [{\citenamefont {Shiokawa}\ and\ \citenamefont
  {Hu}(1995)}]{shiokawa1995decoherence}%
  \BibitemOpen
  \bibfield  {author} {\bibinfo {author} {\bibfnamefont {K.}~\bibnamefont
  {Shiokawa}}\ and\ \bibinfo {author} {\bibfnamefont {B.}~\bibnamefont {Hu}},\
  }\bibfield  {title} {\enquote {\bibinfo {title} {Decoherence, delocalization,
  and irreversibility in quantum chaotic systems},}\ }\href@noop {} {\bibfield
  {journal} {\bibinfo  {journal} {Physical Review E}\ }\textbf {\bibinfo
  {volume} {52}},\ \bibinfo {pages} {2497} (\bibinfo {year}
  {1995})}\BibitemShut {NoStop}%
\bibitem [{\citenamefont {Bellomo}, \citenamefont {Franco},\ and\ \citenamefont
  {Compagno}(2007)}]{bellomo2007non}%
  \BibitemOpen
  \bibfield  {author} {\bibinfo {author} {\bibfnamefont {B.}~\bibnamefont
  {Bellomo}}, \bibinfo {author} {\bibfnamefont {R.~L.}\ \bibnamefont {Franco}},
  \ and\ \bibinfo {author} {\bibfnamefont {G.}~\bibnamefont {Compagno}},\
  }\bibfield  {title} {\enquote {\bibinfo {title} {{Non-Markovian} effects on
  the dynamics of entanglement},}\ }\href@noop {} {\bibfield  {journal}
  {\bibinfo  {journal} {Physical Review Letters}\ }\textbf {\bibinfo {volume}
  {99}},\ \bibinfo {pages} {160502} (\bibinfo {year} {2007})}\BibitemShut
  {NoStop}%
\bibitem [{\citenamefont {Thorwart}\ \emph {et~al.}(2009)\citenamefont
  {Thorwart}, \citenamefont {Eckel}, \citenamefont {Reina}, \citenamefont
  {Nalbach},\ and\ \citenamefont {Weiss}}]{thorwart2009enhanced}%
  \BibitemOpen
  \bibfield  {author} {\bibinfo {author} {\bibfnamefont {M.}~\bibnamefont
  {Thorwart}}, \bibinfo {author} {\bibfnamefont {J.}~\bibnamefont {Eckel}},
  \bibinfo {author} {\bibfnamefont {J.~H.}\ \bibnamefont {Reina}}, \bibinfo
  {author} {\bibfnamefont {P.}~\bibnamefont {Nalbach}}, \ and\ \bibinfo
  {author} {\bibfnamefont {S.}~\bibnamefont {Weiss}},\ }\bibfield  {title}
  {\enquote {\bibinfo {title} {Enhanced quantum entanglement in the
  {non-Markovian} dynamics of biomolecular excitons},}\ }\href@noop {}
  {\bibfield  {journal} {\bibinfo  {journal} {Chemical Physics Letters}\
  }\textbf {\bibinfo {volume} {478}},\ \bibinfo {pages} {234--237} (\bibinfo
  {year} {2009})}\BibitemShut {NoStop}%
\bibitem [{\citenamefont {de~Vega}\ and\ \citenamefont
  {Alonso}(2017)}]{de_vega_dynamics_2017}%
  \BibitemOpen
  \bibfield  {author} {\bibinfo {author} {\bibfnamefont {I.}~\bibnamefont
  {de~Vega}}\ and\ \bibinfo {author} {\bibfnamefont {D.}~\bibnamefont
  {Alonso}},\ }\bibfield  {title} {\enquote {\bibinfo {title} {Dynamics of
  non-{Markovian} open quantum systems},}\ }\href@noop {} {\bibfield  {journal}
  {\bibinfo  {journal} {Reviews of Modern Physics}\ }\textbf {\bibinfo {volume}
  {89}} (\bibinfo {year} {2017})}\BibitemShut {NoStop}%
\bibitem [{\citenamefont {Nakajima}(1958)}]{nakajima1958quantum}%
  \BibitemOpen
  \bibfield  {author} {\bibinfo {author} {\bibfnamefont {S.}~\bibnamefont
  {Nakajima}},\ }\bibfield  {title} {\enquote {\bibinfo {title} {On quantum
  theory of transport phenomena: steady diffusion},}\ }\href@noop {} {\bibfield
   {journal} {\bibinfo  {journal} {Progress of Theoretical Physics}\ }\textbf
  {\bibinfo {volume} {20}},\ \bibinfo {pages} {948--959} (\bibinfo {year}
  {1958})}\BibitemShut {NoStop}%
\bibitem [{\citenamefont {Zwanzig}(1960)}]{zwanzig1960ensemble}%
  \BibitemOpen
  \bibfield  {author} {\bibinfo {author} {\bibfnamefont {R.}~\bibnamefont
  {Zwanzig}},\ }\bibfield  {title} {\enquote {\bibinfo {title} {Ensemble method
  in the theory of irreversibility},}\ }\href@noop {} {\bibfield  {journal}
  {\bibinfo  {journal} {The Journal of Chemical Physics}\ }\textbf {\bibinfo
  {volume} {33}},\ \bibinfo {pages} {1338--1341} (\bibinfo {year}
  {1960})}\BibitemShut {NoStop}%
\bibitem [{\citenamefont {Esposito}\ and\ \citenamefont
  {Gaspard}(2003)}]{esposito2003quantum}%
  \BibitemOpen
  \bibfield  {author} {\bibinfo {author} {\bibfnamefont {M.}~\bibnamefont
  {Esposito}}\ and\ \bibinfo {author} {\bibfnamefont {P.}~\bibnamefont
  {Gaspard}},\ }\bibfield  {title} {\enquote {\bibinfo {title} {Quantum master
  equation for a system influencing its environment},}\ }\href@noop {}
  {\bibfield  {journal} {\bibinfo  {journal} {Physical Review E}\ }\textbf
  {\bibinfo {volume} {68}},\ \bibinfo {pages} {066112} (\bibinfo {year}
  {2003})}\BibitemShut {NoStop}%
\bibitem [{\citenamefont {Cao}\ and\ \citenamefont
  {Lu}(2017)}]{cao_lindblad_2017}%
  \BibitemOpen
  \bibfield  {author} {\bibinfo {author} {\bibfnamefont {Y.}~\bibnamefont
  {Cao}}\ and\ \bibinfo {author} {\bibfnamefont {J.}~\bibnamefont {Lu}},\
  }\bibfield  {title} {\enquote {\bibinfo {title} {Lindblad equation and its
  semiclassical limit of the {Anderson}-{Holstein} model},}\ }\href@noop {}
  {\bibfield  {journal} {\bibinfo  {journal} {Journal of Mathematical Physics}\
  }\textbf {\bibinfo {volume} {58}},\ \bibinfo {pages} {122105} (\bibinfo
  {year} {2017})}\BibitemShut {NoStop}%
\bibitem [{\citenamefont {Van~Kampen}(1992)}]{van1992stochastic}%
  \BibitemOpen
  \bibfield  {author} {\bibinfo {author} {\bibfnamefont {N.~G.}\ \bibnamefont
  {Van~Kampen}},\ }\href@noop {} {\emph {\bibinfo {title} {Stochastic processes
  in Physics and chemistry}}},\ Vol.~\bibinfo {volume} {1}\ (\bibinfo
  {publisher} {Elsevier},\ \bibinfo {year} {1992})\BibitemShut {NoStop}%
\bibitem [{\citenamefont {Shi}\ and\ \citenamefont
  {Geva}(2003)}]{shi_new_2003}%
  \BibitemOpen
  \bibfield  {author} {\bibinfo {author} {\bibfnamefont {Q.}~\bibnamefont
  {Shi}}\ and\ \bibinfo {author} {\bibfnamefont {E.}~\bibnamefont {Geva}},\
  }\bibfield  {title} {\enquote {\bibinfo {title} {A new approach to
  calculating the memory kernel of the generalized quantum master equation for
  an arbitrary system–bath coupling},}\ }\href@noop {} {\bibfield  {journal}
  {\bibinfo  {journal} {The Journal of Chemical Physics}\ }\textbf {\bibinfo
  {volume} {119}},\ \bibinfo {pages} {12063--12076} (\bibinfo {year}
  {2003})}\BibitemShut {NoStop}%
\bibitem [{\citenamefont {Lindblad}(1976)}]{lindblad1976generators}%
  \BibitemOpen
  \bibfield  {author} {\bibinfo {author} {\bibfnamefont {G.}~\bibnamefont
  {Lindblad}},\ }\bibfield  {title} {\enquote {\bibinfo {title} {On the
  generators of quantum dynamical semigroups},}\ }\href@noop {} {\bibfield
  {journal} {\bibinfo  {journal} {Communications in Mathematical Physics}\
  }\textbf {\bibinfo {volume} {48}},\ \bibinfo {pages} {119--130} (\bibinfo
  {year} {1976})}\BibitemShut {NoStop}%
\bibitem [{\citenamefont {Pfalzgraff}\ \emph {et~al.}(2019)\citenamefont
  {Pfalzgraff}, \citenamefont {Montoya-Castillo}, \citenamefont {Kelly},\ and\
  \citenamefont {Markland}}]{pfalzgraff_efficient_2019}%
  \BibitemOpen
  \bibfield  {author} {\bibinfo {author} {\bibfnamefont {W.~C.}\ \bibnamefont
  {Pfalzgraff}}, \bibinfo {author} {\bibfnamefont {A.}~\bibnamefont
  {Montoya-Castillo}}, \bibinfo {author} {\bibfnamefont {A.}~\bibnamefont
  {Kelly}}, \ and\ \bibinfo {author} {\bibfnamefont {T.~E.}\ \bibnamefont
  {Markland}},\ }\bibfield  {title} {\enquote {\bibinfo {title} {Efficient
  construction of generalized master equation memory kernels for multi-state
  systems from nonadiabatic quantum-classical dynamics},}\ }\href@noop {}
  {\bibfield  {journal} {\bibinfo  {journal} {The Journal of Chemical Physics}\
  }\textbf {\bibinfo {volume} {150}},\ \bibinfo {pages} {244109} (\bibinfo
  {year} {2019})}\BibitemShut {NoStop}%
\bibitem [{\citenamefont {Meier}\ and\ \citenamefont
  {Tannor}(1999)}]{meier_non-markovian_1999}%
  \BibitemOpen
  \bibfield  {author} {\bibinfo {author} {\bibfnamefont {C.}~\bibnamefont
  {Meier}}\ and\ \bibinfo {author} {\bibfnamefont {D.~J.}\ \bibnamefont
  {Tannor}},\ }\bibfield  {title} {\enquote {\bibinfo {title} {Non-{Markovian}
  evolution of the density operator in the presence of strong laser fields},}\
  }\href@noop {} {\bibfield  {journal} {\bibinfo  {journal} {The Journal of
  Chemical Physics}\ }\textbf {\bibinfo {volume} {111}},\ \bibinfo {pages}
  {3365--3376} (\bibinfo {year} {1999})}\BibitemShut {NoStop}%
\bibitem [{\citenamefont {Mori}(1965)}]{Mori65}%
  \BibitemOpen
  \bibfield  {author} {\bibinfo {author} {\bibfnamefont {H.}~\bibnamefont
  {Mori}},\ }\bibfield  {title} {\enquote {\bibinfo {title} {Transport,
  collective motion, and {Brownian} motion},}\ }\href@noop {} {\bibfield
  {journal} {\bibinfo  {journal} {Prog. Theor. Phys.}\ }\textbf {\bibinfo
  {volume} {33}},\ \bibinfo {pages} {423 -- 450} (\bibinfo {year}
  {1965})}\BibitemShut {NoStop}%
\bibitem [{\citenamefont {Kelly}\ \emph {et~al.}(2016)\citenamefont {Kelly},
  \citenamefont {Montoya-Castillo}, \citenamefont {Wang},\ and\ \citenamefont
  {Markland}}]{kelly2016generalized}%
  \BibitemOpen
  \bibfield  {author} {\bibinfo {author} {\bibfnamefont {A.}~\bibnamefont
  {Kelly}}, \bibinfo {author} {\bibfnamefont {A.}~\bibnamefont
  {Montoya-Castillo}}, \bibinfo {author} {\bibfnamefont {L.}~\bibnamefont
  {Wang}}, \ and\ \bibinfo {author} {\bibfnamefont {T.~E.}\ \bibnamefont
  {Markland}},\ }\bibfield  {title} {\enquote {\bibinfo {title} {Generalized
  quantum master equations in and out of equilibrium: When can one win?}}\
  }\href@noop {} {\bibfield  {journal} {\bibinfo  {journal} {The Journal of
  Chemical Physics}\ }\textbf {\bibinfo {volume} {144}},\ \bibinfo {pages}
  {184105} (\bibinfo {year} {2016})}\BibitemShut {NoStop}%
\bibitem [{\citenamefont {Montoya-Castillo}\ and\ \citenamefont
  {Reichman}(2017)}]{montoya2017approximate}%
  \BibitemOpen
  \bibfield  {author} {\bibinfo {author} {\bibfnamefont {A.}~\bibnamefont
  {Montoya-Castillo}}\ and\ \bibinfo {author} {\bibfnamefont {D.~R.}\
  \bibnamefont {Reichman}},\ }\bibfield  {title} {\enquote {\bibinfo {title}
  {Approximate but accurate quantum dynamics from the {Mori} formalism. ii.
  equilibrium time correlation functions},}\ }\href@noop {} {\bibfield
  {journal} {\bibinfo  {journal} {The Journal of Chemical Physics}\ }\textbf
  {\bibinfo {volume} {146}},\ \bibinfo {pages} {084110} (\bibinfo {year}
  {2017})}\BibitemShut {NoStop}%
\bibitem [{\citenamefont {Montoya-Castillo}\ and\ \citenamefont
  {Reichman}(2016)}]{montoya2016approximate}%
  \BibitemOpen
  \bibfield  {author} {\bibinfo {author} {\bibfnamefont {A.}~\bibnamefont
  {Montoya-Castillo}}\ and\ \bibinfo {author} {\bibfnamefont {D.~R.}\
  \bibnamefont {Reichman}},\ }\bibfield  {title} {\enquote {\bibinfo {title}
  {Approximate but accurate quantum dynamics from the {Mori} formalism: I.
  nonequilibrium dynamics},}\ }\href@noop {} {\bibfield  {journal} {\bibinfo
  {journal} {The Journal of Chemical Physics}\ }\textbf {\bibinfo {volume}
  {144}},\ \bibinfo {pages} {184104} (\bibinfo {year} {2016})}\BibitemShut
  {NoStop}%
\bibitem [{\citenamefont {Chru\'{s}ci\'{s}ki}\ and\ \citenamefont
  {Kossakowski}(2010)}]{chruscinski_non-markovian_2010}%
  \BibitemOpen
  \bibfield  {author} {\bibinfo {author} {\bibfnamefont {D.}~\bibnamefont
  {Chru\'{s}ci\'{s}ki}}\ and\ \bibinfo {author} {\bibfnamefont
  {A.}~\bibnamefont {Kossakowski}},\ }\bibfield  {title} {\enquote {\bibinfo
  {title} {Non-{Markovian} {Quantum} {Dynamics}: {Local} versus {Nonlocal}},}\
  }\href@noop {} {\bibfield  {journal} {\bibinfo  {journal} {Physical Review
  Letters}\ }\textbf {\bibinfo {volume} {104}},\ \bibinfo {pages} {070406}
  (\bibinfo {year} {2010})}\BibitemShut {NoStop}%
\bibitem [{\citenamefont {Banchi}\ \emph {et~al.}(2018)\citenamefont {Banchi},
  \citenamefont {Grant}, \citenamefont {Rocchetto},\ and\ \citenamefont
  {Severini}}]{banchi2018modelling}%
  \BibitemOpen
  \bibfield  {author} {\bibinfo {author} {\bibfnamefont {L.}~\bibnamefont
  {Banchi}}, \bibinfo {author} {\bibfnamefont {E.}~\bibnamefont {Grant}},
  \bibinfo {author} {\bibfnamefont {A.}~\bibnamefont {Rocchetto}}, \ and\
  \bibinfo {author} {\bibfnamefont {S.}~\bibnamefont {Severini}},\ }\bibfield
  {title} {\enquote {\bibinfo {title} {Modelling {non--markovian} quantum
  processes with recurrent neural networks},}\ }\href@noop {} {\bibfield
  {journal} {\bibinfo  {journal} {New Journal of Physics}\ }\textbf {\bibinfo
  {volume} {20}},\ \bibinfo {pages} {123030} (\bibinfo {year}
  {2018})}\BibitemShut {NoStop}%
\bibitem [{\citenamefont {de~Vega}\ \emph {et~al.}(2005)\citenamefont
  {de~Vega}, \citenamefont {Alonso}, \citenamefont {Gaspard},\ and\
  \citenamefont {Strunz}}]{de_vega_non-markovian_2005}%
  \BibitemOpen
  \bibfield  {author} {\bibinfo {author} {\bibfnamefont {I.}~\bibnamefont
  {de~Vega}}, \bibinfo {author} {\bibfnamefont {D.}~\bibnamefont {Alonso}},
  \bibinfo {author} {\bibfnamefont {P.}~\bibnamefont {Gaspard}}, \ and\
  \bibinfo {author} {\bibfnamefont {W.~T.}\ \bibnamefont {Strunz}},\ }\bibfield
   {title} {\enquote {\bibinfo {title} {Non-{Markovian} stochastic
  {Schrödinger} equations in different temperature regimes: {A} study of the
  spin-boson model},}\ }\href@noop {} {\bibfield  {journal} {\bibinfo
  {journal} {The Journal of Chemical Physics}\ }\textbf {\bibinfo {volume}
  {122}},\ \bibinfo {pages} {124106} (\bibinfo {year} {2005})}\BibitemShut
  {NoStop}%
\bibitem [{\citenamefont {Gaspard}\ and\ \citenamefont
  {Nagaoka}(1999{\natexlab{a}})}]{gaspard1999non}%
  \BibitemOpen
  \bibfield  {author} {\bibinfo {author} {\bibfnamefont {P.}~\bibnamefont
  {Gaspard}}\ and\ \bibinfo {author} {\bibfnamefont {M.}~\bibnamefont
  {Nagaoka}},\ }\bibfield  {title} {\enquote {\bibinfo {title} {Non-markovian
  stochastic {S}chr{\"o}dinger equation},}\ }\href@noop {} {\bibfield
  {journal} {\bibinfo  {journal} {The Journal of Chemical Physics}\ }\textbf
  {\bibinfo {volume} {111}},\ \bibinfo {pages} {5676--5690} (\bibinfo {year}
  {1999}{\natexlab{a}})}\BibitemShut {NoStop}%
\bibitem [{\citenamefont {Hortikar}\ and\ \citenamefont
  {Srednicki}(1998)}]{hortikar1998correlations}%
  \BibitemOpen
  \bibfield  {author} {\bibinfo {author} {\bibfnamefont {S.}~\bibnamefont
  {Hortikar}}\ and\ \bibinfo {author} {\bibfnamefont {M.}~\bibnamefont
  {Srednicki}},\ }\bibfield  {title} {\enquote {\bibinfo {title} {Correlations
  in chaotic eigenfunctions at large separation},}\ }\href@noop {} {\bibfield
  {journal} {\bibinfo  {journal} {Physical review letters}\ }\textbf {\bibinfo
  {volume} {80}},\ \bibinfo {pages} {1646} (\bibinfo {year}
  {1998})}\BibitemShut {NoStop}%
\bibitem [{\citenamefont {Srednicki}(1994)}]{srednicki1994chaos}%
  \BibitemOpen
  \bibfield  {author} {\bibinfo {author} {\bibfnamefont {M.}~\bibnamefont
  {Srednicki}},\ }\bibfield  {title} {\enquote {\bibinfo {title} {Chaos and
  quantum thermalization},}\ }\href@noop {} {\bibfield  {journal} {\bibinfo
  {journal} {Physical Review E}\ }\textbf {\bibinfo {volume} {50}},\ \bibinfo
  {pages} {888} (\bibinfo {year} {1994})}\BibitemShut {NoStop}%
\bibitem [{\citenamefont {Di{\'o}si}(1996)}]{diosi1996exact}%
  \BibitemOpen
  \bibfield  {author} {\bibinfo {author} {\bibfnamefont {L.}~\bibnamefont
  {Di{\'o}si}},\ }\bibfield  {title} {\enquote {\bibinfo {title} {Exact
  semiclassical wave equation for stochastic quantum optics},}\ }\href@noop {}
  {\bibfield  {journal} {\bibinfo  {journal} {Quantum and Semiclassical Optics:
  Journal of the European Optical Society Part B}\ }\textbf {\bibinfo {volume}
  {8}},\ \bibinfo {pages} {309} (\bibinfo {year} {1996})}\BibitemShut {NoStop}%
\bibitem [{\citenamefont {Di{\'o}si}\ and\ \citenamefont
  {Strunz}(1997)}]{diosi1997non}%
  \BibitemOpen
  \bibfield  {author} {\bibinfo {author} {\bibfnamefont {L.}~\bibnamefont
  {Di{\'o}si}}\ and\ \bibinfo {author} {\bibfnamefont {W.~T.}\ \bibnamefont
  {Strunz}},\ }\bibfield  {title} {\enquote {\bibinfo {title} {The
  non-markovian stochastic schr{\"o}dinger equation for open systems},}\
  }\href@noop {} {\bibfield  {journal} {\bibinfo  {journal} {Physics Letters
  A}\ }\textbf {\bibinfo {volume} {235}},\ \bibinfo {pages} {569--573}
  (\bibinfo {year} {1997})}\BibitemShut {NoStop}%
\bibitem [{\citenamefont {Strunz}(1996)}]{strunz1996linear}%
  \BibitemOpen
  \bibfield  {author} {\bibinfo {author} {\bibfnamefont {W.~T.}\ \bibnamefont
  {Strunz}},\ }\bibfield  {title} {\enquote {\bibinfo {title} {Linear quantum
  state diffusion for {non--Markovian} open quantum systems},}\ }\href@noop {}
  {\bibfield  {journal} {\bibinfo  {journal} {Physics Letters A}\ }\textbf
  {\bibinfo {volume} {224}},\ \bibinfo {pages} {25--30} (\bibinfo {year}
  {1996})}\BibitemShut {NoStop}%
\bibitem [{\citenamefont {Suess}, \citenamefont {Eisfeld},\ and\ \citenamefont
  {Strunz}(2014)}]{suess_hierarchy_2014}%
  \BibitemOpen
  \bibfield  {author} {\bibinfo {author} {\bibfnamefont {D.}~\bibnamefont
  {Suess}}, \bibinfo {author} {\bibfnamefont {A.}~\bibnamefont {Eisfeld}}, \
  and\ \bibinfo {author} {\bibfnamefont {W.}~\bibnamefont {Strunz}},\
  }\bibfield  {title} {\enquote {\bibinfo {title} {Hierarchy of {Stochastic}
  {Pure} {States} for {Open} {Quantum} {System} {Dynamics}},}\ }\href@noop {}
  {\bibfield  {journal} {\bibinfo  {journal} {Physical Review Letters}\
  }\textbf {\bibinfo {volume} {113}} (\bibinfo {year} {2014})}\BibitemShut
  {NoStop}%
\bibitem [{\citenamefont {Ke}\ and\ \citenamefont
  {Zhao}(2016)}]{ke2016hierarchy}%
  \BibitemOpen
  \bibfield  {author} {\bibinfo {author} {\bibfnamefont {Y.}~\bibnamefont
  {Ke}}\ and\ \bibinfo {author} {\bibfnamefont {Y.}~\bibnamefont {Zhao}},\
  }\bibfield  {title} {\enquote {\bibinfo {title} {Hierarchy of
  forward--backward stochastic {Schr\"{o}dinger} equation},}\ }\href@noop {}
  {\bibfield  {journal} {\bibinfo  {journal} {The Journal of Chemical Physics}\
  }\textbf {\bibinfo {volume} {145}},\ \bibinfo {pages} {024101} (\bibinfo
  {year} {2016})}\BibitemShut {NoStop}%
\bibitem [{\citenamefont {Gisin}\ and\ \citenamefont
  {Percival}(1992)}]{gisin1992quantum}%
  \BibitemOpen
  \bibfield  {author} {\bibinfo {author} {\bibfnamefont {N.}~\bibnamefont
  {Gisin}}\ and\ \bibinfo {author} {\bibfnamefont {I.~C.}\ \bibnamefont
  {Percival}},\ }\bibfield  {title} {\enquote {\bibinfo {title} {The
  quantum-state diffusion model applied to open systems},}\ }\href@noop {}
  {\bibfield  {journal} {\bibinfo  {journal} {Journal of Physics A:
  Mathematical and General}\ }\textbf {\bibinfo {volume} {25}},\ \bibinfo
  {pages} {5677} (\bibinfo {year} {1992})}\BibitemShut {NoStop}%
\bibitem [{\citenamefont {Gaspard}\ and\ \citenamefont
  {Nagaoka}(1999{\natexlab{b}})}]{gaspard1999slippage}%
  \BibitemOpen
  \bibfield  {author} {\bibinfo {author} {\bibfnamefont {P.}~\bibnamefont
  {Gaspard}}\ and\ \bibinfo {author} {\bibfnamefont {M.}~\bibnamefont
  {Nagaoka}},\ }\bibfield  {title} {\enquote {\bibinfo {title} {Slippage of
  initial conditions for the {Redfield} master equation},}\ }\href@noop {}
  {\bibfield  {journal} {\bibinfo  {journal} {The Journal of Chemical Physics}\
  }\textbf {\bibinfo {volume} {111}},\ \bibinfo {pages} {5668--5675} (\bibinfo
  {year} {1999}{\natexlab{b}})}\BibitemShut {NoStop}%
\bibitem [{\citenamefont {Strunz}, \citenamefont {Diósi},\ and\ \citenamefont
  {Gisin}(1999)}]{strunz_open_1999}%
  \BibitemOpen
  \bibfield  {author} {\bibinfo {author} {\bibfnamefont {W.~T.}\ \bibnamefont
  {Strunz}}, \bibinfo {author} {\bibfnamefont {L.}~\bibnamefont {Diósi}}, \
  and\ \bibinfo {author} {\bibfnamefont {N.}~\bibnamefont {Gisin}},\ }\bibfield
   {title} {\enquote {\bibinfo {title} {Open {System} {Dynamics} with
  {Non}--{Markovian} {Quantum} {Trajectories}},}\ }\href@noop {} {\bibfield
  {journal} {\bibinfo  {journal} {Phys. Rev. Lett.}\ }\textbf {\bibinfo
  {volume} {82}},\ \bibinfo {pages} {1801--1805} (\bibinfo {year}
  {1999})}\BibitemShut {NoStop}%
\bibitem [{\citenamefont {Kubo}(1966)}]{Kubo66}%
  \BibitemOpen
  \bibfield  {author} {\bibinfo {author} {\bibfnamefont {R.}~\bibnamefont
  {Kubo}},\ }\bibfield  {title} {\enquote {\bibinfo {title} {The
  fluctuation-dissipation theorem},}\ }\href@noop {} {\bibfield  {journal}
  {\bibinfo  {journal} {Rep. Prog. Phys.}\ }\textbf {\bibinfo {volume}
  {29(1)}},\ \bibinfo {pages} {255 -- 284} (\bibinfo {year}
  {1966})}\BibitemShut {NoStop}%
\bibitem [{\citenamefont {Tanimura}(2006)}]{tanimura2006stochastic}%
  \BibitemOpen
  \bibfield  {author} {\bibinfo {author} {\bibfnamefont {Y.}~\bibnamefont
  {Tanimura}},\ }\bibfield  {title} {\enquote {\bibinfo {title} {Stochastic
  {Liouville, Langevin, Fokker--Planck}, and master equation approaches to
  quantum dissipative systems},}\ }\href@noop {} {\bibfield  {journal}
  {\bibinfo  {journal} {Journal of the Physical Society of Japan}\ }\textbf
  {\bibinfo {volume} {75}},\ \bibinfo {pages} {082001} (\bibinfo {year}
  {2006})}\BibitemShut {NoStop}%
\bibitem [{\citenamefont {Doob}(1944)}]{Doob44}%
  \BibitemOpen
  \bibfield  {author} {\bibinfo {author} {\bibfnamefont {J.~L.}\ \bibnamefont
  {Doob}},\ }\bibfield  {title} {\enquote {\bibinfo {title} {The elementary
  {Gaussian} processes},}\ }\href@noop {} {\bibfield  {journal} {\bibinfo
  {journal} {Ann. Math. Stat.}\ }\textbf {\bibinfo {volume} {15}},\ \bibinfo
  {pages} {229--282} (\bibinfo {year} {1944})}\BibitemShut {NoStop}%
\bibitem [{\citenamefont {{\O}ksendal}(2003)}]{oksendal2003stochastic}%
  \BibitemOpen
  \bibfield  {author} {\bibinfo {author} {\bibfnamefont {B.}~\bibnamefont
  {{\O}ksendal}},\ }\href@noop {} {\emph {\bibinfo {title} {Stochastic
  differential equations}}}\ (\bibinfo  {publisher} {Springer},\ \bibinfo
  {year} {2003})\BibitemShut {NoStop}%
\bibitem [{\citenamefont {Kloeden}\ and\ \citenamefont
  {Platen}(2013)}]{kloeden2013numerical}%
  \BibitemOpen
  \bibfield  {author} {\bibinfo {author} {\bibfnamefont {P.~E.}\ \bibnamefont
  {Kloeden}}\ and\ \bibinfo {author} {\bibfnamefont {E.}~\bibnamefont
  {Platen}},\ }\href@noop {} {\emph {\bibinfo {title} {Numerical solution of
  stochastic differential equations}}},\ Vol.~\bibinfo {volume} {23}\ (\bibinfo
   {publisher} {Springer Science \& Business Media},\ \bibinfo {year}
  {2013})\BibitemShut {NoStop}%
\bibitem [{\citenamefont {Iacus}(2009)}]{iacus2009simulation}%
  \BibitemOpen
  \bibfield  {author} {\bibinfo {author} {\bibfnamefont {S.~M.}\ \bibnamefont
  {Iacus}},\ }\href@noop {} {\emph {\bibinfo {title} {Simulation and inference
  for stochastic differential equations: with R examples}}}\ (\bibinfo
  {publisher} {Springer Science \& Business Media},\ \bibinfo {year}
  {2009})\BibitemShut {NoStop}%
\bibitem [{\citenamefont {Biele}\ and\ \citenamefont
  {D’Agosta}(2012)}]{biele2012stochastic}%
  \BibitemOpen
  \bibfield  {author} {\bibinfo {author} {\bibfnamefont {R.}~\bibnamefont
  {Biele}}\ and\ \bibinfo {author} {\bibfnamefont {R.}~\bibnamefont
  {D’Agosta}},\ }\bibfield  {title} {\enquote {\bibinfo {title} {A stochastic
  approach to open quantum systems},}\ }\href@noop {} {\bibfield  {journal}
  {\bibinfo  {journal} {Journal of Physics: Condensed Matter}\ }\textbf
  {\bibinfo {volume} {24}},\ \bibinfo {pages} {273201} (\bibinfo {year}
  {2012})}\BibitemShut {NoStop}%
\bibitem [{\citenamefont {Di~Ventra}\ and\ \citenamefont
  {D’Agosta}(2007)}]{di2007stochastic}%
  \BibitemOpen
  \bibfield  {author} {\bibinfo {author} {\bibfnamefont {M.}~\bibnamefont
  {Di~Ventra}}\ and\ \bibinfo {author} {\bibfnamefont {R.}~\bibnamefont
  {D’Agosta}},\ }\bibfield  {title} {\enquote {\bibinfo {title} {Stochastic
  time-dependent current-density-functional theory},}\ }\href@noop {}
  {\bibfield  {journal} {\bibinfo  {journal} {Physical Review etters}\ }\textbf
  {\bibinfo {volume} {98}},\ \bibinfo {pages} {226403} (\bibinfo {year}
  {2007})}\BibitemShut {NoStop}%
\bibitem [{\citenamefont {Risken}(1984)}]{risken1984fokker}%
  \BibitemOpen
  \bibfield  {author} {\bibinfo {author} {\bibfnamefont {H.}~\bibnamefont
  {Risken}},\ }\href@noop {} {\emph {\bibinfo {title} {{F}okker-{P}lanck
  Equation}}}\ (\bibinfo  {publisher} {Springer},\ \bibinfo {year}
  {1984})\BibitemShut {NoStop}%
\bibitem [{\citenamefont {Ritschel}\ and\ \citenamefont
  {Eisfeld}(2014)}]{ritschel2014analytic}%
  \BibitemOpen
  \bibfield  {author} {\bibinfo {author} {\bibfnamefont {G.}~\bibnamefont
  {Ritschel}}\ and\ \bibinfo {author} {\bibfnamefont {A.}~\bibnamefont
  {Eisfeld}},\ }\bibfield  {title} {\enquote {\bibinfo {title} {Analytic
  representations of bath correlation functions for ohmic and superohmic
  spectral densities using simple poles},}\ }\href@noop {} {\bibfield
  {journal} {\bibinfo  {journal} {The Journal of Chemical Physics}\ }\textbf
  {\bibinfo {volume} {141}},\ \bibinfo {pages} {094101} (\bibinfo {year}
  {2014})}\BibitemShut {NoStop}%
\bibitem [{\citenamefont {Li}\ and\ \citenamefont
  {Li}(2020)}]{li2020exponential}%
  \BibitemOpen
  \bibfield  {author} {\bibinfo {author} {\bibfnamefont {J.}~\bibnamefont
  {Li}}\ and\ \bibinfo {author} {\bibfnamefont {X.}~\bibnamefont {Li}},\
  }\bibfield  {title} {\enquote {\bibinfo {title} {Exponential integrators for
  stochastic {Schr{\"o}dinger} equations},}\ }\href@noop {} {\bibfield
  {journal} {\bibinfo  {journal} {Physical Review E}\ }\textbf {\bibinfo
  {volume} {101}},\ \bibinfo {pages} {013312} (\bibinfo {year}
  {2020})}\BibitemShut {NoStop}%
\bibitem [{\citenamefont {Suzuki}\ and\ \citenamefont
  {Yamauchi}(1993)}]{suzuki1993convergence}%
  \BibitemOpen
  \bibfield  {author} {\bibinfo {author} {\bibfnamefont {M.}~\bibnamefont
  {Suzuki}}\ and\ \bibinfo {author} {\bibfnamefont {T.}~\bibnamefont
  {Yamauchi}},\ }\bibfield  {title} {\enquote {\bibinfo {title} {Convergence of
  unitary and complex decompositions of exponential operators},}\ }\href@noop
  {} {\bibfield  {journal} {\bibinfo  {journal} {Journal of mathematical
  Physics}\ }\textbf {\bibinfo {volume} {34}},\ \bibinfo {pages} {4892--4897}
  (\bibinfo {year} {1993})}\BibitemShut {NoStop}%
\bibitem [{\citenamefont {Telatovich}\ and\ \citenamefont
  {Li}(2017)}]{telatovich2017strong}%
  \BibitemOpen
  \bibfield  {author} {\bibinfo {author} {\bibfnamefont {A.}~\bibnamefont
  {Telatovich}}\ and\ \bibinfo {author} {\bibfnamefont {X.}~\bibnamefont
  {Li}},\ }\bibfield  {title} {\enquote {\bibinfo {title} {The strong
  convergence of operator-splitting methods for the {Langevin} dynamics
  model},}\ }\href@noop {} {\bibfield  {journal} {\bibinfo  {journal} {arXiv
  preprint arXiv:1706.04237}\ } (\bibinfo {year} {2017})}\BibitemShut {NoStop}%
\bibitem [{\citenamefont {Hochbruck}\ and\ \citenamefont
  {Lubich}(1997)}]{hochbruck1997krylov}%
  \BibitemOpen
  \bibfield  {author} {\bibinfo {author} {\bibfnamefont {M.}~\bibnamefont
  {Hochbruck}}\ and\ \bibinfo {author} {\bibfnamefont {C.}~\bibnamefont
  {Lubich}},\ }\bibfield  {title} {\enquote {\bibinfo {title} {On {Krylov}
  subspace approximations to the matrix exponential operator},}\ }\href@noop {}
  {\bibfield  {journal} {\bibinfo  {journal} {SIAM Journal on Numerical
  Analysis}\ }\textbf {\bibinfo {volume} {34}},\ \bibinfo {pages} {1911--1925}
  (\bibinfo {year} {1997})}\BibitemShut {NoStop}%
\bibitem [{\citenamefont {Castro}, \citenamefont {Marques},\ and\ \citenamefont
  {Rubio}(2004)}]{castro2004propagators}%
  \BibitemOpen
  \bibfield  {author} {\bibinfo {author} {\bibfnamefont {A.}~\bibnamefont
  {Castro}}, \bibinfo {author} {\bibfnamefont {M.~A.}\ \bibnamefont {Marques}},
  \ and\ \bibinfo {author} {\bibfnamefont {A.}~\bibnamefont {Rubio}},\
  }\bibfield  {title} {\enquote {\bibinfo {title} {Propagators for the
  time-dependent {Kohn--Sham} equations},}\ }\href@noop {} {\bibfield
  {journal} {\bibinfo  {journal} {The Journal of Chemical Physics}\ }\textbf
  {\bibinfo {volume} {121}},\ \bibinfo {pages} {3425--3433} (\bibinfo {year}
  {2004})}\BibitemShut {NoStop}%
\bibitem [{\citenamefont {Ma}, \citenamefont {Li},\ and\ \citenamefont
  {Liu}(2017)}]{ma2017fluctuation}%
  \BibitemOpen
  \bibfield  {author} {\bibinfo {author} {\bibfnamefont {L.}~\bibnamefont
  {Ma}}, \bibinfo {author} {\bibfnamefont {X.}~\bibnamefont {Li}}, \ and\
  \bibinfo {author} {\bibfnamefont {C.}~\bibnamefont {Liu}},\ }\bibfield
  {title} {\enquote {\bibinfo {title} {Fluctuation-dissipation theorem
  consistent approximation of the {Langevin} dynamics model},}\ }\href@noop {}
  {\bibfield  {journal} {\bibinfo  {journal} {Communications in Mathematical
  Sciences}\ }\textbf {\bibinfo {volume} {15}},\ \bibinfo {pages} {1171--1181}
  (\bibinfo {year} {2017})}\BibitemShut {NoStop}%
\bibitem [{\citenamefont {Yoshida}(1990)}]{yoshida1990construction}%
  \BibitemOpen
  \bibfield  {author} {\bibinfo {author} {\bibfnamefont {H.}~\bibnamefont
  {Yoshida}},\ }\bibfield  {title} {\enquote {\bibinfo {title} {Construction of
  higher order symplectic integrators},}\ }\href@noop {} {\bibfield  {journal}
  {\bibinfo  {journal} {Physics letters A}\ }\textbf {\bibinfo {volume}
  {150}},\ \bibinfo {pages} {262--268} (\bibinfo {year} {1990})}\BibitemShut
  {NoStop}%
\bibitem [{\citenamefont {Chorin}\ and\ \citenamefont
  {Lu}(2015)}]{chorin2015discrete}%
  \BibitemOpen
  \bibfield  {author} {\bibinfo {author} {\bibfnamefont {A.~J.}\ \bibnamefont
  {Chorin}}\ and\ \bibinfo {author} {\bibfnamefont {F.}~\bibnamefont {Lu}},\
  }\bibfield  {title} {\enquote {\bibinfo {title} {Discrete approach to
  stochastic parametrization and dimension reduction in nonlinear dynamics},}\
  }\href@noop {} {\bibfield  {journal} {\bibinfo  {journal} {Proceedings of the
  National Academy of Sciences}\ }\textbf {\bibinfo {volume} {112}},\ \bibinfo
  {pages} {9804--9809} (\bibinfo {year} {2015})}\BibitemShut {NoStop}%
\bibitem [{\citenamefont {Lei}, \citenamefont {Baker},\ and\ \citenamefont
  {Li}(2016)}]{lei2016generalized}%
  \BibitemOpen
  \bibfield  {author} {\bibinfo {author} {\bibfnamefont {H.}~\bibnamefont
  {Lei}}, \bibinfo {author} {\bibfnamefont {N.}~\bibnamefont {Baker}}, \ and\
  \bibinfo {author} {\bibfnamefont {X.}~\bibnamefont {Li}},\ }\bibfield
  {title} {\enquote {\bibinfo {title} {The generalized {Langevin} equation and
  the parameterization from data},}\ }\href@noop {} {\bibfield  {journal}
  {\bibinfo  {journal} {Proc. Natl. Acad. Sci.}\ }\textbf {\bibinfo {volume}
  {In press}} (\bibinfo {year} {2016})}\BibitemShut {NoStop}%
\bibitem [{\citenamefont {Bishwal}(2007)}]{bishwal2007parameter}%
  \BibitemOpen
  \bibfield  {author} {\bibinfo {author} {\bibfnamefont {J.~P.}\ \bibnamefont
  {Bishwal}},\ }\href@noop {} {\emph {\bibinfo {title} {Parameter estimation in
  stochastic differential equations}}}\ (\bibinfo  {publisher} {Springer},\
  \bibinfo {year} {2007})\BibitemShut {NoStop}%
\bibitem [{\citenamefont {Song}\ and\ \citenamefont
  {Park}(2016)}]{song2016linear}%
  \BibitemOpen
  \bibfield  {author} {\bibinfo {author} {\bibfnamefont {X.}~\bibnamefont
  {Song}}\ and\ \bibinfo {author} {\bibfnamefont {J.~H.}\ \bibnamefont
  {Park}},\ }\bibfield  {title} {\enquote {\bibinfo {title} {Linear optimal
  estimation for discrete-time measurement delay systems with multichannel
  multiplicative noise},}\ }\href@noop {} {\bibfield  {journal} {\bibinfo
  {journal} {IEEE Transactions on Circuits and Systems II: Express Briefs}\
  }\textbf {\bibinfo {volume} {64}},\ \bibinfo {pages} {156--160} (\bibinfo
  {year} {2016})}\BibitemShut {NoStop}%
\bibitem [{\citenamefont {Pavliotis}(2014)}]{Pav_book:14}%
  \BibitemOpen
  \bibfield  {author} {\bibinfo {author} {\bibfnamefont {G.~A.}\ \bibnamefont
  {Pavliotis}},\ }\href@noop {} {\emph {\bibinfo {title} {Stochastic processes
  and applications: diffusion processes, the {Fokker-Planck and Langevin}
  equations}}},\ Vol.~\bibinfo {volume} {60}\ (\bibinfo  {publisher}
  {Springer},\ \bibinfo {year} {2014})\BibitemShut {NoStop}%
\end{thebibliography}%

\end{document}